\documentclass[a4paper,12pt]{article}
\pdfoutput=1 
\usepackage{jcappub} 
\usepackage[T1]{fontenc} 
\title{Electroweak Vacuum Instability and Renormalized Higgs Field Vacuum Fluctuations in the Inflationary Universe}
\author[a,b]{Kazunori Kohri,}
\author[b]{Hiroki Matsui}
\affiliation[a]{Institute of Particle and Nuclear Studies, KEK,\\ 1-1 Oho, Tsukuba 305-0801, Japan}
\affiliation[b]{The Graduate University for Advanced Studies (SOKENDAI),\\ 1-1 Oho, Tsukuba 305-0801, Japan}
\emailAdd{kohri@post.kek.jp}
\emailAdd{matshiro@post.kek.jp}

\abstract{
In this work, we investigated the electroweak vacuum instability during or after inflation.
In the inflationary Universe, i.e., de Sitter space, the vacuum field fluctuations $\left< {\delta \phi  }^{ 2 } \right>$ 
enlarge in proportion to the Hubble scale $H^{2}$. 
Therefore, the large inflationary vacuum fluctuations of the Higgs field $\left< {\delta \phi  }^{ 2 } \right>$ 
are potentially catastrophic to trigger the vacuum transition to the
negative-energy Planck-scale vacuum state and cause an immediate collapse of the Universe. 
However, the vacuum field fluctuations $\left< {\delta \phi  }^{ 2 } \right>$, i.e., 
the vacuum expectation values have an ultraviolet divergence, and therefore a renormalization is necessary 
to estimate the physical effects of the vacuum transition. 
Thus, in this paper, we revisit the electroweak vacuum instability from the perspective of 
quantum field theory (QFT) in curved space-time, and discuss the dynamical behavior of the homogeneous Higgs field $\phi$ 
determined by the effective potential ${ V }_{\rm eff}\left( \phi \right)$ in curved space-time
and the renormalized vacuum fluctuations $\left< {\delta \phi  }^{ 2 } \right>_{\rm ren}$
via adiabatic regularization and point-splitting regularization.
We simply suppose that the Higgs field only couples the gravity via the
non-minimal Higgs-gravity coupling $\xi(\mu)$.
In this scenario, the electroweak vacuum stability is inevitably threatened by the dynamical behavior of 
the homogeneous Higgs field  $\phi$, or the formations of AdS domains or bubbles
unless the Hubble scale is small enough $H< \Lambda_{I} $.
}

\begin{document}
\maketitle
\flushbottom
\section{Introduction}
\label{sec:intro}

The recent measurements of the Higgs boson mass $m_{h}=125.09\ \pm \  0.21\ ({\rm stat})\ \pm \  0.11\ ({\rm syst})\ {\rm GeV}$~\cite{Aad:2015zhl,Aad:2013wqa,Chatrchyan:2013mxa,Giardino:2013bma} and the top quark mass 
$m_{t}=173.34\pm 0.27\ ({\rm stat})\ {\rm GeV}$~\cite{2014arXiv1403.4427A} suggest 
that the current electroweak vacuum state of the Universe is not stable, and finally cause
a catastrophic vacuum decay through quantum tunneling~\cite{Kobzarev:1974cp,Coleman:1977py,Callan:1977pt}
although the cosmological timescale for the quantum tunneling decay is longer than the age of the Universe
~\cite{Degrassi:2012ry,Isidori:2001bm,Ellis:2009tp,EliasMiro:2011aa}. 
In de Sitter space, especially the inflationary Universe, however, the curved background enlarges the vacuum field fluctuations 
$\left<{  \delta \phi   }^{ 2 } \right>$ in proportion to the Hubble scale $H^{2}$.
Therefore, if the large inflationary vacuum fluctuations $\left<{  \delta \phi   }^{ 2 } \right>$ of the Higgs field overcomes the  barrier of the standard model Higgs effective potential $V_{\rm eff}\left( \phi  \right)$, it triggers off a catastrophic 
vacuum transition to the negative Planck-energy true vacuum and cause an immediate collapse of the Universe
~\cite{Espinosa:2007qp,Fairbairn:2014zia,Lebedev:2012sy,Kobakhidze:2013tn,Enqvist:2013kaa,Herranen:2014cua,Kobakhidze:2014xda,Kamada:2014ufa,Enqvist:2014bua,Hook:2014uia,Kearney:2015vba,Espinosa:2015qea,Kohri:2016wof,East:2016anr}.

The vacuum field fluctuations $\left<{  \delta \phi   }^{ 2 } \right>$, i.e., the vacuum expectation values are formally  given by  
\begin{eqnarray}
\left< { \delta \phi^{2}  } \right>=\int { { d }^{ 3 }k{ \left| \delta { \phi  }_{ k }\left( \eta ,x \right)  \right|  }^{ 2 } }
=\int _{ 0 }^{ \infty  }{ \frac { dk }{ k }  } { \Delta   }_{ \delta \phi  }^{2}\left(\eta ,k \right)\label{eq:hfhfhedg} .
\end{eqnarray}
where $ { \Delta   }_{ \delta \phi  }^{2}\left(\eta ,k \right)$ is defined as the power spectrum of quantum vacuum fluctuations.
As well-known facts in quantum field theory (QFT), the vacuum expectation values $\left<{  \delta \phi   }^{ 2 } \right>$
have an ultraviolet divergence (quadratic or logarithmic) and therefore a regularization is necessary. 
The quadratic divergence corresponds to the normal contribution from the fluctuations of the vacuum
in Minkowski space, and it can be eliminated by standard renormalization in flat spacetime. 
The logarithmic divergence, however, appears as a consequence of the expansion of the Universe, 
and has the physical contributions to the origin of the primordial perturbations or the backreaction of the inflaton field.
We usually eliminate this logarithmic ultraviolet divergence 
by simply neglecting the modes with $k > aH$.
That corresponds to the stochastic Fokker-Planck (FP) equation, which treats
the inflationary field fluctuations generally originating from long wave modes, i.e., the IR parts~\cite{Linde:1993xx}.
Recent works~\cite{Hook:2014uia,Kearney:2015vba,Espinosa:2015qea,Kohri:2016wof,East:2016anr}
for the electroweak vacuum stability during inflation based on the stochastic Fokker-Planck (FP) equation.
However, from the viewpoint of QFT, 
we must treat carefully short wave modes as well as long wave modes~\cite{Parker:2007ni,Agullo:2011qg} and 
it is necessary to renormalize the vacuum expectation values $\left<{  \delta \phi   }^{ 2 } \right>$
in the curved space-time in order to obtain exact physical contributions~\cite{Vilenkin:1982wt,Paz:1988mt}.

Thus, in this paper, we revisit the electroweak vacuum instability from the legitimate perspective of QFT in curved space-time.
In the first part of this paper, we derive the one-loop Higgs effective potential in curved space-time 
via the adiabatic expansion method.
In the second part, we discuss the renormalized field vacuum fluctuations $\left<{  \delta \phi   }^{ 2 } \right>$ 
in de Sitter space by using adiabatic regularization and point-splitting regularization. 
In the third part,  we investigate the electroweak vacuum instability during or after inflation
from the global and homogeneous Higgs field $\phi$, and 
the renormalized vacuum fluctuations  $\left<{  \delta \phi   }^{ 2 } \right>_{\rm ren}$
corresponding to the local and inhomogeneous Higgs field fluctuations.
The behavior of the homogeneous Higgs field $\phi$ is determined by 
the effective potential $V_{\rm eff}\left( \phi  \right)$ in curved space-time, and then,
the excursion of the homogeneous Higgs field $\phi$
to the negative Planck-energy vacuum state can terminate inflation and triggers off a catastrophic collapse of the Universe.
The local and inhomogeneous Higgs fluctuations described by $\left<{  \delta \phi   }^{ 2 } \right>_{\rm ren}$
generate catastrophic Anti-de Sitter (AdS) domains or bubbles and finally cause a vacuum transition.
In this work, we improve our previous work~\cite{Kohri:2016wof},
provide a comprehensive study of the phenomenon and reach new conclusions.
In addition, we persist in the zero-temperature field theory, 
leaving the generalization to the finite-temperature case and discussion of 
the thermal History of the metastable Universe for a forthcoming work.

This paper is organized as follows. In Section~\ref{sec:effective} 
we derive the Higgs effective potential in curved space-time by using the adiabatic expansion method. 
In Section~\ref{sec:adiabatic} we discuss the problem of renormalization
to the vacuum fluctuations in de Sitter space by using adiabatic regularization. 
In Section~\ref{sec:point} we consider the renormalized vacuum fluctuations via point-splitting regularization
and show that the renormalized expectation values via point-splitting regularization is consistent with 
the previous results via adiabatic regularization.
In Section~\ref{sec:electroweak} we discuss the behavior of the global Higgs field $\phi$ and 
the vacuum transitions via the renormalized Higgs field vacuum fluctuations $\left<{  \delta \phi   }^{ 2 } \right>_{\rm ren}$,
and investigate the electroweak vacuum instability during inflation or after inflation. 
Finally, in Section~\ref{sec:Conclusion} we draw the conclusion of our work.

\section{Higgs effective potential in curved space-time via adiabatic expansion method}
\label{sec:effective}
The behavior of the homogeneous Higgs field $\phi$ on the entire Universe is determined 
by the effective potential $V_{\rm eff}\left( \phi  \right)$ in curved space-time.
In this section, we review the standard derivation of the one-loop effective potential in curved space-time 
via the adiabatic expansion method~\cite{Maroto:2014oca,Albareti:2016cvx,Ringwald:1987ui,Sinha:1988ci,Huang:1993fk}, 
and then, show how the effective potential $V_{\rm eff}\left( \phi  \right)$ of curved space-time
is changed from the Minkowski space-time, where we use the notations and the conventions of 
Ref.\cite{Maroto:2014oca,Albareti:2016cvx}. 
For simplicity, we consider a flat Robertson-Walker background not including metric perturbations.
Thus, the Friedmann-Lemaitre-Robertson-Walker (FLRW) metric is given by
\begin{equation}
g_{\mu\nu}={\rm diag}\left( -1,\frac { { a }^{ 2 }\left( t \right)  }{ 1-K { r }^{ 2 } } ,{ a }^{ 2 }\left( t \right) { r }^{ 2 },{ a }^{ 2 }\left( t \right) { r }^{ 2 }\sin^{2} { \theta  }  \right)\label{eqsdffedg},
\end{equation}
where $a = a\left(t\right)$ is the scale factor with the cosmic time $t$ and $K $ is the curvature parameter, 
where positive, zero, and negative values correspond to closed, flat, and hyperbolic space-time, respectively.
For the spatially flat Universe, we take $K=0$.  Then, the scalar curvature is obtained as
\begin{equation}
R=6\left[ 
{ \left( \frac { \dot { a }  }{ a }  \right)  }^{ 2 }+\left( \frac { \ddot { a }  }{ a }  \right)  
\right]=6\left(\frac{a''}{a^{3}}\right)\label{eq:dgdgfdgedg}. 
\end{equation}
where the conformal time $\eta$ has been introduced and is defined by $d\eta=dt/a$.
In the de Sitter space-time, the scale factor becomes $a\left(t\right) =e^{Ht}$ or  $a\left(\eta\right) =-1/H\eta$, 
the scalar curvature is estimated to be $R=12H^{2}$ in the de Sitter Universe.

The action for the Higgs field with the potential $V\left( \phi  \right)$ in curved space-time is given by
\begin{equation}
S\left[ \phi  \right] =-\int { { d }^{ 4 }x\sqrt { -g } \left( \frac { 1 }{ 2 } { g }^{ \mu \nu  }
{ \nabla  }_{ \mu  }\phi {\nabla  }_{ \nu  }\phi +V\left( \phi  \right)  \right)  } \label{eq:ddddddgedg},
\end{equation}
where we assume the simple form for the Higgs potential as
\begin{equation}
V\left( \phi  \right) =\frac{1}{2}\left(m^{2}+\xi R\right)\phi^{2}+\frac{\lambda}{4}\phi^{4} \label{eq:aaaadg}.
\end{equation}
Thus, the Klein-Gordon equation for the Higgs field are given by
\begin{equation}
\Box \phi\left(\eta ,x\right) +V'\left( \phi\left(\eta ,x\right)  \right) =0 \label{eq:dsssssdg},
\end{equation}
where $\Box$ denotes the generally covariant d'Alembertian operator, 
$\Box =g^{\mu\nu}{ \nabla  }_{ \mu  }{  \nabla   }_{ \nu  }=1/\sqrt { -g } { \partial  }_{ \mu  }\left( \sqrt { -g } { \partial  }^{ \mu  } \right) $
and $\xi$ is the non-minimal Higgs-gravity coupling constant. 
There are two popular choices for $\xi$, i.e., minimal coupling ($\xi=0$) and conformal coupling ($\xi=1/6$),
which is conformally invariant in the massless limit.
However, the non-minimal Higgs-gravity coupling $\xi$ is inevitably generated through the loop corrections.

In the quantum field theory, we treat the Higgs field $\phi \left(\eta ,x\right)$ as the operator acting on the states. 
We assume that the vacuum expectation value of the Higgs field is 
$\phi\left(\eta \right)=\left< 0  \right| { \phi \left(\eta ,x\right) }\left| 0 \right>$. 
In this case, the Higgs field  $\phi \left(\eta ,x\right)$ 
can be decomposed into a classic component and a quantum component as
\begin{equation}
\phi \left(\eta ,x\right)=\phi \left(\eta \right)+\delta \phi \left(\eta ,x \right) \label{eq:kkkkgedg},
\end{equation}
where $\left< 0  \right| { \delta\phi \left(\eta ,x\right) }\left| 0 \right>=0$.
In the one-loop approximation, we can obtain the following equations
\begin{eqnarray}
&&\Box \phi +V'\left( \phi  \right) +\frac { 1 }{ 2 } V'''\left( \phi  \right) \left< \delta { \phi  }^{ 2 } \right>  =0 \label{eq:lklkedg},\\
&&\Box \delta \phi +V''\left( \phi  \right) \delta \phi =0 \label{eq:dppppg},
\end{eqnarray}
where the mass of the quantum field $\delta \phi$ is written by
\begin{equation}
V''\left( \phi  \right)={ m }^{ 2 }+3\lambda \phi^{2}+\xi R \label{eq:dyyytyg}.
\end{equation}
The quantum field $\delta \phi$ is decomposed into each $k$ mode,
\begin{equation}
\delta \phi \left( \eta ,x \right) =\int { { d }^{ 3 }k\left( { a }_{ k }\delta { \phi  }_{ k }\left( \eta ,x \right) +{ a }_{ k }^{ \dagger  }\delta { \phi  }_{ k }^{ * }\left( \eta ,x \right)  \right)  }  \label{eq:ddfkkfledg},
\end{equation}
where
\begin{equation}
{ \delta \phi  }_{ k }\left( \eta ,x \right) =\frac { { e }^{ ik\cdot x } }{ { \left( 2\pi  \right)  }^{ 3/2 }
\sqrt { C\left( \eta  \right)  }  } \delta { \chi  }_{ k }\left( \eta  \right)  \label{eq:slkdlkgdg},
\end{equation}
with $C\left(\eta \right)=a^{2}\left(\eta \right)$.
Thus, the Higgs field vacuum fluctuations $\left<{  \delta \phi   }^{ 2 } \right>$, 
i.e., the expectation field values can be written as
\begin{eqnarray}
\left< 0  \right| { \delta \phi^{2}  }\left| 0 \right>&=&\int { { d }^{ 3 }k{ \left| \delta { \phi  }_{ k }\left( \eta ,x \right)  \right|  }^{ 2 } } \label{eq:ddkg;ergedg},\\
&=&\frac { 1 }{ 2{ \pi  }^{ 2 }C\left( \eta  \right)  } \int _{ 0 }^{ \infty  } { dk { k }^{ 2 }{ \left| \delta { \chi  }_{ k } \right|  }^{ 2 } }  \label{eq:xkkgdddgedg},
\end{eqnarray}
where $\left<{  \delta \phi   }^{ 2 } \right>$ has an ultraviolet divergence (quadratic and logarithmic) 
and requires a regularization, e.g. cut-off regularization or dimensional regularization. 
From Eq.~(\ref{eq:dppppg}), the Klein-Gordon equation for the quantum field $\delta \chi$ is written by
\begin{equation}
{ \delta \chi}''_{ k }+{ \Omega  }_{ k }^{ 2 }\left( \eta  \right) { \delta \chi  }_{ k }=0 \label{eq:dlkrlekg}.
\end{equation}
Here, we use the adiabatic (WKB) approximation to obtain the mode function. 
For the lowest-order approximation,  
the time-dependent mode function is given by
\begin{equation}
\delta{ \chi  }_{ k }=\frac { 1 }{ \sqrt { 2{\Omega_{k}\left( \eta  \right)   }}  } 
\exp\left( -i\int { \Omega_{k} \left( \eta  \right) d\eta  }  \right)  \label{eq:ddkreitjdg},
\end{equation}
where ${ \Omega  }_{ k }^{ 2 }\left( \eta  \right) ={ k }^{ 2 }+C\left( \eta  \right) \left( { m }^{ 2 } 
+3\lambda \phi^{2}+\left( \xi -1/6 \right) R \right) $.
More precisely, we must consider the higher order approximation and include the exact effects of the particle 
productions for the one-loop effective potential (see Ref.\cite{Ringwald:1987ui} for the details).
However, we can simply include such effects by adding the backreaction term of 
the vacuum filed fluctuations, i.e., $\left<{  \delta \phi   }^{ 2 } \right>_{\rm ren}$.
Furthermore, we comment the condition of the adiabatic (WKB) approximation (${ \Omega  }_{ k }^{ 2 }>0$ and 
$\left| { \Omega ' }_{ k }/{ \Omega  }_{ k }^{ 2 } \right| \ll 1$). This condition breaks during  
inflation for the massless scalar field, or in the parametric resonance of the preheating (see, e.g. Ref.\cite{Kofman:1997yn}).
In these cases, the IR parts at $k<aH$  breaks the WKB approximation and we can expect enormous particle production.
However, the UV parts at $k>aH$ are not affected by the cosmological dynamics of the Universe,
and the effective potential generally originates from the UV parts, i.e., short wave modes.
Therefore, as well as the radiation dominated and the matter dominated eras, 
we can adopt the adiabatic expansion method for the effective potential 
in de Sitter space by taking into account the IR backreaction effects (see Ref.\cite{Ringwald:1987ui} for the details).
As a consequence, we must add the backreaction term, i.e., renormalized field vacuum fluctuations 
$\left<{  \delta \phi   }^{ 2 } \right>_{\rm ren}$ (see Section~\ref{sec:adiabatic} for the details)
to the effective potential in curved space-time.
Then, we write the expectation field value as follows 
\begin{equation}
\left< 0  \right| { \delta \phi^{2}  }\left| 0 \right>
=\frac { 1 }{ 4{ \pi  }^{ 2 }a^{2} } \int_{ 0 }^{ \infty  }{ dk\frac { { k }^{ 2 } }{ \sqrt { { k }^{ 2 }+
\left( { m }^{ 2 }+3\lambda \phi^{2}+\left( \xi -1/6 \right) R \right) a^{2}  }  }  }  \label{eq:dklgkhldg}.
\end{equation}
The one-loop contribution to the Higgs effective potential is given by
\begin{eqnarray}
&&\frac { 1 }{ 2 } V'''\left( \phi  \right) \left< 0  \right| { \delta \phi^{2}  }\left| 0 \right> \nonumber \\
&=&\frac{d}{d\phi}\left(\frac { 1 }{ 4{ \pi  }^{ 2 }a^{4} } \int _{ }^{ \Lambda  }{ dk\ { k }^{ 2 }\sqrt { { k }^{ 2 }+\left( { m }^{ 2 }+3\lambda \phi^{2}+\left( \xi -1/6 \right) R \right) a^{2}  }  }\right), \nonumber \\
&=&\frac{dV_{1}\left( \phi  \right)}{d\phi} \label{eq:dskjksjg},
\end{eqnarray}
where we take an ultraviolet cut-off as $\Lambda$ in order to regularize quadratic or logarithmic divergence.
For convenience, we rewrite the classic Higgs field equation as follows
\begin{eqnarray}
\Box \phi +V'\left( \phi  \right)+ V'_{1}\left( \phi  \right) =0 \label{eq:dklhkgkdg}.
\end{eqnarray}
In order to obtain the effective potential, we calculate exactly the integral
\begin{eqnarray}
V_{1}\left( \phi  \right) &=&\frac { 1 }{ 4{ \pi  }^{ 2 }a^{4} } \int _{ }^{ \Lambda  }
{ dk\ { k }^{ 2 }\sqrt { { k }^{ 2 }+\left( { m }^{ 2 }+3\lambda \phi^{2}+\left( \xi -1/6 \right) R \right) a^{2}  }  } ,\\
&=& \frac { 1 }{ 32{ \pi  }^{ 2 }a^{4} } \Biggl[ \left( \Lambda \left( 2{ \Lambda  }^{ 2 }+\left( { m }^{ 2 }+3\lambda \phi^{2}+\left( \xi -1/6 \right) R \right) a^{2} \right)  \right) \nonumber  \\ && \times \sqrt { { \Lambda  }^{ 2 }+\left( { m }^{ 2 }+3\lambda \phi^{2}
+\left( \xi -1/6 \right) R \right) a^{2} } 
+\left( { m }^{ 2 }+3\lambda \phi^{2}+\left( \xi -1/6 \right) R \right)^{2}a^{4} \nonumber  \\ &&
 \times \ln { \left( \frac {\left( { m }^{ 2 }+3\lambda \phi^{2}+\left( \xi -1/6 \right) R \right)^{1/2}a }{ \Lambda +\sqrt { { \Lambda  }^{ 2 }+\left( { m }^{ 2 }+3\lambda \phi^{2}+\left( \xi -1/6 \right) R \right) a^{2}  }  }  \right)  }   \Biggr]  \label{ezzddgedg}.
\end{eqnarray}
In the limit $\Lambda \rightarrow \infty$, we can obtain the following expression 
\begin{eqnarray}
V_{1}\left( \phi  \right) &=&
\frac { { \Lambda  }^{ 4 } }{ 16{ \pi  }^{ 2 }{ a }^{ 4 } } +\frac { { \left( { m }^{ 2 }+3\lambda \phi^{2}+\left( \xi -1/6 \right) R \right)\Lambda  }^{ 2 } }{ 16{ \pi  }^{ 2 }{ a }^{ 2 } }
-\frac { { \left( { m }^{ 2 }+3\lambda \phi^{2}+\left( \xi -1/6 \right) R \right)}^{ 2 } }{ 64{ \pi  }^{ 2 } } \ln { \left( \frac { { \Lambda  }^{ 2 } }{ { \mu  }^{ 2 } }  \right)  } \nonumber \\&&
+\frac { { \left( { m }^{ 2 }+3\lambda \phi^{2}+\left( \xi -1/6 \right) R \right) }^{ 2 } }{ 64{ \pi  }^{ 2 } } \ln { \left( \frac { \left( { m }^{ 2 }+3\lambda \phi^{2}+\left( \xi -1/6 \right) R \right){ a }^{ 2 } }{ { \mu  }^{ 2 } }  \right)  } \nonumber \\ && 
+\frac { { \left( { m }^{ 2 }+3\lambda \phi^{2}+\left( \xi -1/6 \right) R \right) }^{ 2 } }{ 64{ \pi  }^{ 2 } }C_{i}
\label{eq:ddjkjkjdf},
\end{eqnarray}
where we introduced the renormalization scale $\mu$ and the constant $C_{i}= \left(1/2-2\ln { 2 }\right)$ depends on the regularization method and the renormalization scheme. Here, we focus on the divergent contribution to the effective potential, which is given by
\begin{eqnarray}
V_{\Lambda}\left( \phi  \right) &&=
\frac { { \Lambda  }^{ 4 } }{ 16{ \pi  }^{ 2 }{ a }^{ 4 } } +\frac { { \left( { m }^{ 2 }+3\lambda \phi^{2}+\left( \xi -1/6 \right) R \right)\Lambda  }^{ 2 } }{ 16{ \pi  }^{ 2 }{ a }^{ 2 } } 
-\frac { { \left( { m }^{ 2 }+\left( \xi -1/6 \right) R \right)}^{ 2 } }{ 64{ \pi  }^{ 2 } } \ln { \left( \frac { { \Lambda  }^{ 2 } }{ { \mu  }^{ 2 } }  \right)  } \nonumber \\ && 
+\left( \frac { 3\lambda { \Lambda  }^{ 2 } }{ 16{ \pi  }^{ 2 }a^{2}} -\frac { 6\lambda{ \left( { m }^{ 2 }+\left( \xi -1/6 \right) R \right)} }{ 64{ \pi  }^{ 2 } } \ln { \left( \frac { { \Lambda  }^{ 2 } }{ { \mu  }^{ 2 } }  \right)  }  \right) { \phi  }^{ 2 }-\frac { 9{ \lambda  }^{ 2 } }{ 64{ \pi  }^{ 2 } } \ln { \left( \frac { { \Lambda  }^{ 2 } }{ { \mu  }^{ 2 } }  \right) { \phi  }^{ 4 } }  \label{eq:asmsmedg}.
\end{eqnarray}
For convenience, we replace $\Lambda \rightarrow a\Lambda$, $\mu/a \rightarrow \mu$ 
and the divergent contribution can be written as
\begin{eqnarray}
V_{\Lambda }\left( \phi  \right) &&=
\frac { { \Lambda  }^{ 4 } }{ 16{ \pi  }^{ 2 } } +\frac { { \left( { m }^{ 2 }+3\lambda \phi^{2}+\left( \xi -1/6 \right) R \right)\Lambda  }^{ 2 } }{ 16{ \pi  }^{ 2 } } -\frac { { \left( { m }^{ 2 }+\left( \xi -1/6 \right) R \right)}^{ 2 } }{ 64{ \pi  }^{ 2 } } \ln { \left( \frac {{ \Lambda  }^{ 2 } }{ { \mu  }^{ 2 } }  \right)  } \nonumber \\ && +\left( \frac { 3\lambda { \Lambda  }^{ 2 } }{ 16{ \pi  }^{ 2 } } -\frac { 6\lambda{ \left( { m }^{ 2 }+\left( \xi -1/6 \right) R \right)}}{ 64{ \pi  }^{ 2 } } \ln { \left( \frac {{ \Lambda  }^{ 2 } }{ { \mu  }^{ 2 } }  \right)  }  \right) { \phi  }^{ 2 }  -\frac { 9{ \lambda  }^{ 2 } }{ 64{ \pi  }^{ 2 } } \ln { \left( \frac {{ \Lambda  }^{ 2 } }{ { \mu  }^{ 2 } }  \right) { \phi  }^{ 4 } }  \label{eq:dklkfdg}.
\end{eqnarray}
We can obviously remove all the divergences (quartic, quadratic and logarithmic)
by absorbing the counter-terms as follows:
\begin{equation}
V_{\rm eff}\left( \phi  \right) =V\left( \phi  \right)+ V_{1}\left( \phi  \right) +
{ \delta  }_{ cc}+\frac { 1 }{ 2 } { \delta  }_{ m }{ \phi  }^{ 2 }+\frac { 1 }{ 2 } { \delta  }_{\xi }{ \phi  }^{ 2 }
+\frac { 1 }{ 4 } { \delta  }_{ \lambda  }{ \phi  }^{ 4 } \label{eq:ddkrekkdg}.
\end{equation}
We obtain the Higgs effective potential in curved space-time as follows: 
\begin{eqnarray}
V_{\rm eff}\left( \phi  \right) &&=\frac{1}{2}m^{2}\phi^{2}+\frac{1}{2}\xi R\phi^{2}
+\frac{\lambda}{4}\phi^{4} \label{eq:jjfkgedg}\\ 
&&+\frac { { \left( { m }^{ 2 }+3\lambda \phi^{2}+\left( \xi -1/6 \right) R \right) }^{ 2 } }{ 64{ \pi  }^{ 2 } }
\ln { \left( \frac { { m }^{ 2 }+3\lambda \phi^{2}+\left( \xi -1/6 \right) R}{ { \mu  }^{ 2 } } -C_{i} \right)  }\nonumber,
\end{eqnarray}
which is consistent with the results by using the heat kernel method~\cite{PhysRevD.31.953,PhysRevD.31.2439}.
The effective potential in curved background has been thoroughly investigated in the literatures~\cite{Herranen:2014cua,
Toms:1983qr,Buchbinder:1985js,Hu:1984js,Balakrishnan:1991pm,Muta:1991mw,Kirsten:1993jn,Elizalde:1993ee,
Elizalde:1993ew,Elizalde:1993qh,Elizalde:1994im,Elizalde:1994ds,Gorbar:2002pw,Gorbar:2003yt,Gorbar:2003yp,
Czerwinska:2015xwa} and 
there are a variety of ways to derive the effective potential in curved spacetime.
Now, we can read off the scale dependence of ${ m }^{ 2 }$, $\xi$ and $\lambda$
from the Eq.~(\ref{eq:jjfkgedg}), and the $\beta$ functions are given by 
\begin{eqnarray}
\beta_{\lambda } &\equiv& \frac { d\lambda  }{ d\ln { \mu  }  } =\frac { 18{ \lambda  }^{ 2 } }{ { \left( 4\pi  \right)  }^{ 2 } }  \label{eq:dkldgedg},\\
\beta_{\xi } &\equiv& \frac { d\xi  }{ d\ln { \mu  }  } =\frac { 6{ \lambda  }  }{ { \left( 4\pi  \right)  }^{ 2 } }\left( \xi -1/6 \right) \label{eq:dgeddlkgg}, \\
\beta_{ m^{2} }&\equiv& \frac { d{ { m }^{ 2 } }  }{ d\ln { \mu  }  } =\frac { 6{ \lambda  }m^{2} }{ { \left( 4\pi  \right)  }^{ 2 } }  \label{eq:ddakdlgg}.
\end{eqnarray}
Finally, we can rewrite the classical Higgs field equation by using the effective potential in curved space-time as follows:
\begin{eqnarray}
\Box \phi + V'_{\rm eff}\left( \phi  \right) =0 \label{eq:ddlreredg}.
\end{eqnarray}
The effective potential given by Eq.~(\ref{eq:jjfkgedg}) does not include the exact particle productions
and we must consider the backreaction term from the renormalized field vacuum fluctuations 
$\left<{  \delta \phi   }^{ 2 } \right>_{\rm ren}$ to improve the effective potential in curved space-time.

\section{Renormalized vacuum fluctuations via adiabatic regularization}
\label{sec:adiabatic}
The main difficulty with the vacuum instability in de Sitter space 
comes from the vacuum field fluctuations on the dynamical background. 
The renormalized vacuum fluctuations $\left< { \delta \phi   }^{ 2 } \right>_{\rm ren}$ originate from
the dynamical particle production effects, and it corresponds to the local and inhomogeneous Higgs field fluctuations. 
In this section, we discuss the renormalized vacuum fluctuations $\left< { \delta \phi   }^{ 2 } \right>_{\rm ren}$ 
by using adiabatic regularization. The adiabatic regularization~\cite{0305-4470-13-4-022,PhysRevD.10.3905,FULLING1974176,PhysRevD.9.341,Haro:2010zz,Haro:2010mx}  is the extremely powerful method to remove the ultraviolet divergence from the expectation field value $\left< { \delta \phi   }^{ 2 } \right>$ and obtain the renormalized  finite value, which has physical contribution. 
For convenience, we write the equation of the field $\delta\chi$ for conformal time coordinate $\eta$ given by
\begin{equation}
{ \delta \chi}''_{ k }+{ \Omega  }_{ k }^{ 2 }\left( \eta  \right) { \delta \chi  }_{ k }=0 \label{eq:dlkrlekg},
\end{equation}
where ${ \Omega  }_{ k }^{ 2 }\left( \eta  \right) ={ \omega  }_{ k }^{ 2 }\left( \eta  \right) + C\left( \eta  \right) \left( \xi -1/6 \right) R $ and ${ \omega  }_{ k }^{ 2 }\left( \eta  \right) ={ k }^{ 2 }+C\left( \eta  \right) \left( { m }^{ 2 }+3\lambda \phi^{2} \right) $.
More precisely, the self-coupling term $3\lambda \phi^{2}$ includes the backreaction effect, i.e., $3\lambda\phi^{2}+3\lambda\left< { \delta \phi   }^{ 2 } \right>_{\rm ren}$ where we shift the Higgs field $\phi^{2}\rightarrow \phi^{2}+\left< {  \delta \phi  }^{ 2 } \right>_{\rm ren} $. Therefore, the vacuum field fluctuations on the dynamical background become complicated and intricate in contrast with
static space-time. 
For simplicity, we neglect the self-coupling term $3\lambda \phi^{2}$ in Section~\ref{sec:adiabatic} and Section~\ref{sec:point}.
The Wronskian condition is given as
\begin{equation}
{ \delta\chi  }_{ k }{ \delta\chi  }_{ k }^{ * }-{ \delta\chi  }_{ k }^{ * }{ \delta\chi  }_{ k }= i\label{eq:dddrrrg},
\end{equation}
which ensures the canonical commutation relations for the field operator $\delta\chi$ below 
\begin{eqnarray}
\bigl[ { a }_{ k },{ a }_{ k' } \bigr] &=&\bigl[{ a }_{ k }^{ \dagger  },{ a }_{ k' }^{ \dagger  } \bigr]=0 \label{eq:dfphipedg},\\ 
\bigl[{ a }_{ k },{ a }_{ k' }^{ \dagger  } \bigr] &=&\delta \left( k-k' \right) \label{eq:oeuoegedg}.
\end{eqnarray}
The adiabatic vacuum $\left| 0 \right>_{\rm A} $ is the vacuum state which is annihilated by all the operators ${ a }_{ k }$
and defined by choosing $\chi_{k}\left(\eta\right)$ to be a positive-frequency WKB mode.
The adiabatic (WKB) approximation to the time-dependent mode function is written by 
\begin{equation}
{\delta \chi  }_{ k }=\frac { 1 }{ \sqrt { 2{ W }_{ k }\left( \eta  \right) }  } 
\exp\left( -i\int { { W }_{ k }\left( \eta  \right) d\eta  }  \right) \label{eq:jgskedg},
\end{equation}
where
\begin{equation}
{ W }_{ k }^{ 2 }={ \Omega  }_{ k }^{ 2 }-\frac { 1 }{ 2 } \frac { { W }''_{ k } }{ { W }_{ k } } 
+\frac { 3 }{ 4 } \frac { { \left( { W }'_{ k } \right)  }^{ 2 } }{ { W }_{ k }^{ 2 } } \label{eq:sklsdkldg}.
\end{equation}
We can obtain the WKB solution by solving Eq.~(\ref{eq:sklsdkldg}) with an iterative procedure 
and the lowest-order WKB solution ${ W }^{0}_{ k }$ is given by
\begin{equation}
\left({ W }^{0}_{ k }\right)^{2}={ \Omega  }_{ k }^{2}\label{eq:flkldg}.
\end{equation}
The first-order WKB solution ${ W }^{1}_{ k }$ is given by
\begin{equation}
\left({ W }^{1}_{ k }\right)^{2}={ \Omega  }_{ k }^{2}-\frac { 1 }{ 2 } \frac { \left({ W}^{0}_{ k }\right)'' }{{ W }^{0}_{ k } } 
+\frac { 3 }{ 4 } \frac {\left({{ W}^{0}_{ k }}'\right)^{2}  }{ \left({ W}^{0}_{ k }\right)^{2} } \label{eq:skdl;lgedg}.
\end{equation}
For the high-order with the nearly conformal case $\xi \simeq1/6$, we can obtain the following expression 
\begin{eqnarray}
{ W }_{ k }&\simeq&{ \omega  }_{ k }+\frac { 3\left( \xi -1/6 \right)  }{ 4{ \omega  }_{ k } } \left( 2D'+{ D }^{ 2 } \right) -\frac { { m }^{ 2 }C }{ 8{ \omega  }_{ k }^{ 3 } } \left( D'+{ D }^{ 2 } \right) +\frac { 5{ m }^{ 4 }{ C }^{ 2 }{ D }^{ 2 } }{ 32{ \omega  }^{ 5 } } \nonumber \\
&&+\frac { { m }^{ 2 }C }{ 32{ \omega  }_{ k }^{ 5 } } \left( D'''+4D'D+3{ D' }^{ 2 }+6D'D^{ 2 }+D^{ 4 } \right) \nonumber \\
&&-\frac { { m }^{ 4 }{ C }^{ 2 } }{ 128{ \omega  }_{ k }^{ 7 } } \left( 28D''D+19{ D' }^{ 2 }+122{ D' }^{ 2 }+47D^{ 4 } \right)  \nonumber \\ 
&&+\frac { { 221m }^{ 6 }{ C }^{ 3 } }{ 256{ \omega  }_{ k }^{ 9 } } \left( D'D^{ 2 }+D^{ 4 } \right) 
-\frac { 1105{ m }^{ 8 }{ C }^{ 4 }{ D }^{ 4 } }{ 2048{ \omega  }_{ k }^{ 11 } }  \nonumber \\
&&-\frac { \left( \xi -1/6 \right)  }{ 8{ \omega  }_{ k }^{ 3 } } \left( 3D'''+3D''D+3{ D' }^{ 2 } \right) \nonumber \\
&&+\left( \xi -1/6 \right) \frac { { m }^{ 2 }C }{ 32{ \omega  }_{ k }^{ 5 } } 
\left( 30D''D+18{ D' }^{ 2 }+57D'D^{ 2 }+9D^{ 4 } \right)  \nonumber \\
&&-\left( \xi -1/6 \right) \frac { { 75m }^{ 4 }{ C }^{ 2 } }{ 128{ \omega  }_{ k }^{ 7 } } \left( 2D'D^{ 2 }+D^{ 4 } \right) \nonumber \\
&&-\frac { { \left( \xi -1/6 \right)  }^{ 2 } }{ 32{ \omega  }_{ k }^{ 3 } } \left( 36{ D' }^{ 2 }+36D'D^{ 2 }+9D^{ 4 } \right) \label{eq:oekgkf},\end{eqnarray}
where $D=C'/C$. The vacuum expectation values ${ \left< { \delta \phi   }^{ 2 } \right>  }_{ \rm ad }$ 
by using the adiabatic vacuum sate $\left| 0 \right>_{\rm A} $
can be written by
\begin{eqnarray}
{ \left< { \delta \phi   }^{ 2 } \right>  }_{ \rm ad }
={ { _{ A }\left< { 0 }|{{  \delta \phi   }^{ 2 }  }|{ 0 } \right>_{A  } } }
=\frac { 1 }{ 4{ \pi  }^{ 2 }C\left( \eta  \right)  } \int _{ 0 }^{ \Lambda  }{ \frac { { k }^{ 2 } }{ { W }_{ k } } dk } \label{eq:dlajshegedg}.\end{eqnarray}
The adiabatic regularization is not the method of regularizing divergent integrals as 
cut-off regularization or dimensional regularization.
Thus, Eq.~(\ref{eq:dlajshegedg}) includes the divergences which need to be removed by these regularization.
However, the divergences in the exact expression ${ \left< { \delta \phi   }^{ 2 } \right>  }$,
which come from the large $k$  modes, are the same as the divergences in the adiabatic expression 
${ \left< { \delta \phi   }^{ 2 } \right>_{\rm ad}   }$.
Thus, we can remove the divergences by subtracting the adiabatic expression ${ \left< { \delta \phi   }^{ 2 } \right>_{\rm ad}  }$ 
from the original expression ${ \left< { \delta \phi   }^{ 2 } \right>  }$ as follows:
\begin{eqnarray}
{ \left< { \delta \phi   }^{ 2 } \right>  }_{ \rm ren }
&=&{ \left< {  \delta \phi   }^{ 2 } \right>  }-{ \left< {  \delta \phi   }^{ 2 } \right>  }_{ \rm ad },\\
&=&\frac { 1 }{ 4{ \pi  }^{ 2 }C\left( \eta  \right)  }  
\Biggl[ \int _{ 0 }^{ \Lambda  }{ 2k^{2}{ \left| \delta { \chi  }_{ k } \right|  }^{ 2 }dk } 
-\int _{ 0 }^{ \Lambda  }{ \frac { { k }^{ 2 } }{ { W }_{ k } } dk } \Biggr] \label{eq:dkhkkegedg}.
 \end{eqnarray}
This method has been shown to be equivalent to the point-splitting regularization~\cite{Birrell513,Anderson:1987yt},
which has been used in a large number of space-time background.

\subsection{Massless minimally coupled cases}
\label{sec:massless}
In this subsection, we discuss the vacuum expectation values in the massless minimally coupled case ($m = 0$ and $\xi = 0$). 
In the massless minimally coupling case, the power spectrum on super-horizon scales is given by
\begin{equation}
 { \Delta   }_{ \delta \phi  }^{2}\left( k \right)  ={ \left( \frac { H }{ 2\pi  }  \right)  }^{ 2 }\label{eq:hlkhggedg},
 \end{equation}
where $ { \Delta   }_{ \delta \phi  }^{2}\left( k \right)  ={ k }^{ 3 }{ \left| \delta { \phi  }_{ k } \right|  }^{ 2 }/2{ \pi  }^{ 2 }$ 
originates from the facts that the inflationary quantum field fluctuations are constant on super-horizon scales.
If we take $aH$ as the UV cut-off  and $H$ as the IR cut-off, the vacuum expectation values 
$\left< { \delta \phi   }^{ 2 } \right>$ 
are simply given by 
\begin{equation}
\left< { \delta \phi   }^{ 2 } \right>=\int _{ H }^{ aH  }{ \frac { dk }{k }  } { \Delta   }_{ \delta \phi  }^{2}\left( k \right)  
=\frac { { H }^{ 3 } }{ { 4{ \pi  }^{ 2 } } } t \label{eq:rjktjgedg}.
\end{equation}
Here, we review the renormalization of the vacuum expectation values $\left< { \delta \phi   }^{ 2 } \right>$ 
with $m = 0$ and $\xi = 0$ by using the adiabatic regularization, 
where we use the result of Ref.\cite{Haro:2010zz,Haro:2010mx},
and show that Eq.~(\ref{eq:rjktjgedg}) is consistent with the results via the adiabatic regularization.

In this case, the mode function $\delta { \chi  }_{ k }\left( \eta  \right) $ can be exactly given by 
\begin{equation}
\delta { \chi  }_{ k }\left( \eta  \right) ={ a }_{ k }\delta { \varphi  }_{ k }\left( \eta  \right) 
+{ b }_{ k }\delta { \varphi }^{*}_{ k }\left( \eta  \right) \label{eq:ohjijgedg},
\end{equation}
where
\begin{equation}
\delta { \varphi  }_{ k }\left( \eta  \right) =\sqrt { \frac { 1 }{ 2k }  } { e }^{ -ik\eta  }\left( 1+\frac { 1 }{ ik\eta  }  \right)\label{eq:fhghgedg}. 
\end{equation}
In the massless minimally coupled case, the vacuum expectation values $\left< { \delta \phi   }^{ 2 } \right>$ 
have not only ultraviolet divergences but also infrared divergences. To avoid the infrared divergences,
we assume that the Universe changes over from the radiation-dominated phase to the de Sitter phase as the following 
\begin{equation}
a\left( \eta  \right) =\begin{cases} 2-\frac { \eta  }{ { \eta  }_{ 0 } } ,\quad \left( \eta <{ \eta  }_{ 0 } \right)  \\ 
\frac { \eta  }{ { \eta  }_{ 0 } } ,\quad\quad\ \left( \eta >{ \eta  }_{ 0 } \right)  \end{cases}\label{eq:fhghedg}
\end{equation}
where ${ \eta  }_{ 0 }=-1/H$ and, during the radiation-dominated Universe $\left( \eta <{ \eta  }_{ 0 } \right)$, 
we choose the mode function
\begin{equation}
\delta { \chi  }_{ k }={ e }^{ -ik\eta  }/\sqrt { 2k }\label{eq:fhfhedg}.
\end{equation}
which is in-vacuum state.
Requiring the matching conditions $\delta { \chi   }_{ k }\left( \eta  \right)$ and $\delta { \chi   }_{ k }'\left( \eta  \right)$
at ${ \eta  }={ \eta  }_{ 0 }$,  we can obtain the coefficients of the mode function
\begin{equation}
{ a }_{ k }=1+\frac { H }{ ik } -\frac { { H }^{ 2 } }{ 2{ k }^{ 2 } } 
,\quad { b }_{ k }={ a }_{ k }+\frac { 2ik }{ 3H } +O\left( \frac { { k }^{ 2 } }{ { H }^{ 2 } }  \right) \label{eq:fkhkokedg}.
\end{equation}
For small $k$ modes in the de Sitter Universe $\left( \eta  >{ \eta  }_{ 0 } \right)$, we have 
\begin{equation}
{ \left| \delta { \chi  }_{ k } \right|  }^{ 2 }=\frac { 1 }{ 2k } 
\left[ { \left( \frac { 2 }{ 3H\eta  } +2+\frac { { H }^{ 2 }{ \eta  }^{ 2 } }{ 6 }  \right)  }^{ 2 }
+O\left( \frac { { k }^{ 2 } }{ { H }^{ 2 } }  \right) +\cdots  \right] \label{eq:joooegedg}.
\end{equation}
Therefore, we obviously have no infrared divergences because of 
$k^{2}{ \left| \delta { \chi  }_{ k } \right|  }^{ 2 }\sim O\left(k\right)$.
For large $k$ modes,  we can obtain the mode function
\begin{equation}
{ \left| \delta { \chi  }_{ k } \right|  }^{ 2 }=\frac { 1 }{ 2k } \left[ 1+\frac { 1 }{ { k }^{ 2 }{ \eta  }^{ 2 } } -\frac { { H }^{ 2 } }{ { k }^{ 2 } } \cos { \left( 2k\left( 1/H+\eta  \right)  \right)  } +O\left( \frac { { H }^{ 3 } }{ { k }^{ 3 } }  \right) +\cdots  \right] \label{eq:ogfhghedg}.
\end{equation}
Here, we consider the following adiabatic (WKB) solution
\begin{eqnarray}
{ W }_{ k }&=&{ \omega  }_{ k }-\frac { 1 }{ 8{ \omega  }_{ k } }
\left( 2D'+{ D }^{ 2 } \right) -\frac { 1 }{ 8 } \frac { { m }^{ 2 }{ C }'' }{ { \omega  }_{ k }^{ 3 } } 
+\frac { 5 }{ 32 } \frac { { { m }^{ 4 }\left( C' \right)  }^{ 2 } }{ { \omega  }_{ k }^{ 5 } } ,\\
&=&{ \omega  }_{ k }-\frac { 1 }{ \eta^{2} { \omega  }_{ k } }
-\frac { 1 }{ 8 } \frac { { m }^{ 2 }{ C }'' }{ { \omega  }_{ k }^{ 3 } } 
+\frac { 5 }{ 32 } \frac { { { m }^{ 4 }\left( C' \right)  }^{ 2 } }{ { \omega  }_{ k }^{ 5 } } \label{eq:ofhfhgedg}.
\end{eqnarray}
We can obtain 
\begin{equation}
\frac{1}{{ W }_{ k }}\simeq\frac{1}{{ \omega  }_{ k }}+\frac { 1 }{ \eta^{2} { \omega  }_{ k }^{3} }
+\frac { 1 }{ 8 } \frac { { m }^{ 2 }{ C }'' }{ { \omega  }_{ k }^{ 5 } } 
-\frac { 5 }{ 32 } \frac { { { m }^{ 4 }\left( C' \right)  }^{ 2 } }{ { \omega  }_{ k }^{ 7 } } \label{eq:fhfhgedg}.
\end{equation}
The condition of the adiabatic (WKB) approximation, i.e., ${ \Omega  }_{ k }^{ 2 }>0$
requires  $k>\sqrt{2}/\left| \eta \right|=\sqrt{2}aH$ as the cut-off of $k$ mode.
Therefore, Eq.~(\ref{eq:dkhkkegedg}) can be given as follows:
\begin{eqnarray}
{ \left< { \delta \phi   }^{ 2 } \right>  }_{ \rm ren }
&=&\lim _{ m\rightarrow 0 }\frac { 1 }{ 4{ \pi  }^{ 2 }C\left( \eta  \right)  }  
\Biggl[ \int _{ 0 }^{ \Lambda  }{ 2k^{2}{ \left| \delta { \chi  }_{ k } \right|  }^{ 2 }dk } 
-\int _{ \sqrt{2}/\left| \eta \right| }^{ \Lambda  }{ \frac { { k }^{ 2 } }{ { W }_{ k } } dk } \Biggr] , \\
&=&\lim _{ m\rightarrow 0 }\frac { 1 }{ 4{ \pi  }^{ 2 }C\left( \eta  \right)  }
 \Biggl[ \int _{ 0 }^{ \Lambda  }{ 2k^{2}{ \left| \delta { \chi  }_{ k } \right|  }^{ 2 }dk } -\int _{\sqrt{2}/\left| \eta \right| }^{ \Lambda  }{ \frac { { k }^{ 2 } }{ { \omega  }_{ k } } dk}
-\int _{ \sqrt{2}/\left| \eta \right| }^{ \Lambda  }{ \frac { { k }^{ 2 } }{\eta^{2} { \omega  }_{ k }^{3} } dk} \nonumber \\
&&+\frac { { m }^{ 2 }C'' }{ 8 } \int _{ \sqrt{2}/\left| \eta \right| }^{ \Lambda  }{ \frac { { k }^{ 2 } }{ { \omega  }_{ k }^{ 5 } } dk
+\frac { { 5m }^{ 4 }{ \left( C' \right)  }^{ 2 } }{ 32 } \int _{ \sqrt{2}/\left| \eta \right| }^{ \Lambda  }
 { \frac { { k }^{ 2 } }{ { \omega  }_{ k }^{ 7 } } dk }  }\Biggr] \label{eq:fhgegedg}.
  \end{eqnarray}
For large $k$ modes, we can subtract the ultraviolet divergences as
\begin{equation}
\lim _{ m\rightarrow 0 }\frac { 1 }{ 4{ \pi  }^{ 2 }C\left( \eta  \right)  } \int _{ \sqrt{2}/\left| \eta \right| }^{ \Lambda  }
{ \left(k-\frac { { k }^{ 2 } }{ { \omega  }_{ k } }\right)dk } =0\label{eq:ohfgghedg}.
\end{equation}
\begin{equation}
\lim _{ m\rightarrow 0 }\frac { 1 }{ 4{ \pi  }^{ 2 } \eta^{2}C\left( \eta  \right)  } \int _{ \sqrt{2}/\left| \eta \right| }^{ \Lambda  }
{ \left(\frac{1}{k}-\frac { { k }^{ 2 } }{ { \omega  }_{ k }^{3} }\right)dk } =0\label{eq:dk;dkggedg}.
\end{equation}
Furthermore, we can eliminate the following divergences
\begin{equation}
\lim _{ m\rightarrow 0 }\frac { { m }^{ 2 }C'' }{ 8 } \int _{ \sqrt{2}/\left| \eta \right| }^{ \Lambda  }
{ \frac { { k }^{ 2 } }{ { \omega  }_{ k }^{ 5 } } dk = \lim _{ m\rightarrow 0 }
\frac { { 5m }^{ 4 }{ \left( C' \right)  }^{ 2 } }{ 32 } \int _{ \sqrt{2}/\left| \eta \right| }^{ \Lambda  }
 { \frac { { k }^{ 2 } }{ { \omega  }_{ k }^{ 7 } } dk }  } =0\label{eq:fhk;hkegedg}.
 \end{equation}
Therefore, we can obtain the following expression 
\begin{eqnarray}
{ \left< { \delta \phi   }^{ 2 } \right>  }_{ \rm ren }
&=&\frac { 1 }{ 2{ \pi  }^{ 2 }C\left( \eta  \right)  } \int _{ 0 }^{ \sqrt{2}/\left| \eta \right|  }
{ k^{2}{ \left| \delta { \chi  }_{ k } \right|  }^{ 2 }dk } \nonumber \\ &&
+\frac { { \eta  }^{ 2 }{ H }^{ 2 } }{ 4{ \pi  }^{ 2 } } \int _{ \sqrt { 2 } /\left| \eta  \right|  }^{ \infty  }
{ \left( -\frac { { H }^{ 2 } }{ { k }^{ 2 } } \cos { \left( 2k\left( 1/H+\eta  \right)  \right)  } +\cdots  \right)  } kdk\label{eq:oehf:lhg}.
 \end{eqnarray}
At the late cosmic-time ($\eta\simeq0$ that is $N_{\rm tot}=Ht\gg1$), we have the following approximation 
\begin{eqnarray}
{ \left< { \delta \phi   }^{ 2 } \right>  }_{ \rm ren }
&\simeq&\frac { { \eta  }^{ 2 }{ H }^{ 2 } }{ 2{ \pi  }^{ 2 } } \int _{ 0 }^{ \sqrt{2}/\left| \eta \right|  }
{ k^{2}{ \left| \delta { \chi  }_{ k } \right|  }^{ 2 }dk }, \nonumber \\ 
&\simeq&\frac { 1 }{ 9{ \pi  }^{ 2 } } \int _{ 0 }^{ H }{ kdk } 
+\frac { { H }^{ 2 } }{ 4{ \pi  }^{ 2 } } \int _{ H }^{ \sqrt { 2 } /\left| \eta  \right|  }{ \frac { 1 }{ k } dk } \label{eq:oflhkegedg},
 \end{eqnarray}
where we approximate ${ \left| \delta { \chi  }_{ k } \right|  }^{ 2 }=2/9{ \eta  }^{ 2 }{ H }^{ 2 }k$ for small $k$ modes
and ${ \left| \delta { \chi  }_{ k } \right|  }^{ 2 }=1/2\eta^{2}k^{3}$ for large $k$ modes. We can finally obtain
\begin{eqnarray}
{ \left< { \delta \phi   }^{ 2 } \right>  }_{ \rm ren }
&\simeq&\frac { { H }^{ 2 } }{ 18{ \pi  }^{ 2 } } +\frac { { H }^{ 2 } }{ 4{ \pi  }^{ 2 } } 
\left( \frac { 1 }{ 2 } \log { 2 } +Ht \right) \simeq \frac { { H }^{ 3 } }{ 4{ \pi  }^{ 2 } } t\label{eq:odhhghedg},
 \end{eqnarray}
which coincides with Eq.~(\ref{eq:rjktjgedg}) via the physical cut-off.

\subsection{Massive non-minimally coupled cases }
\label{sec:massive}
In this subsection, we consider the massive non-minimally coupled case ($m \neq  0$ and $\xi \neq 0$).
At first, we discuss the renormalized vacuum fluctuations for $m\ll H$.\footnote{In de Sitter space,
the scalar curvature becomes $R=12H^{2}$, and therefore the non-minimal coupling $\xi$ provides
the effective mass-term $m^{2}=12H^{2}\xi$.} 
In $m\ll H$ case, the power spectrum on super-horizon scales can be written as
\begin{equation}
 { \Delta   }_{ \delta \phi  }^{2}\left( k \right)   = { \left( \frac { H }{ 2\pi  }  \right)  }^{ 2 }{ \left( \frac { k }{ aH }  \right)  }^{ 3-2\nu  }\label{eq:hfojjgfghg},
 \end{equation}
where  $\nu=\sqrt { 9/4-{ m }^{ 2 }/{ H }^{ 2 } }$.
If we take $aH$ as the UV cut-off  and $H$ as the IR cut-off, 
the vacuum expectation values $\left< { \delta \phi   }^{ 2 } \right>$ can be given by 
\begin{equation}
\left< { \delta \phi   }^{ 2 } \right>=\int _{ H }^{ aH  }{ \frac { dk }{k }  }  { \Delta   }_{ \delta \phi  }^{2}\left( k \right)  
=\frac { 3{ H }^{ 4 } }{ 8{ \pi  }^{ 2 }{ m }^{ 2 } } \label{eq:gmkmkmg}.
\end{equation}
Here, we briefly discuss the renormalized vacuum fluctuations ${ \left< { \delta \phi   }^{ 2 } \right>  }_{ \rm ren }$
for $m\ll H$ by using the adiabatic regularization.
The UV parts can be eliminated by using the adiabatic regularization, i.e., 
the following integral of Eq.~(\ref{eq:dkhkkegedg}) can converge 
\begin{equation}
{ \left< { \delta \phi   }^{ 2 } \right>  }_{ \rm div }
=\frac { 1 }{ 2{ \pi  }^{ 2 }C\left( \eta  \right)  }  \int _{  \sqrt { 2 } /\left| \eta  \right|  }^{\Lambda }
{k^{2}\left( { \left| \delta { \chi  }_{ k } \right|  }^{ 2 }-\frac { 1}{ 2{ W }_{ k } }\right)dk } \label{eq:fhfggedg},
\end{equation}
where we take the mode function corresponding to the exact Bunch-Davies vacuum state given by 
\begin{equation}
\delta { \chi  }_{ k }=\sqrt { \frac { \pi  }{ 4 }  } { \eta  }^{ 1/2 }{ H }_{ \nu  }^{ \left( 1 \right)  }\left( k\eta  \right) \label{eq:hjojojhg},\end{equation}
with ${ H }_{ \nu  }^{ \left( 1 \right)  }$ is the Hankel function of the first kind.
By discarding the convergent terms, whose orders are $O(m^2)$, 
we can obtain the following expression 
\begin{eqnarray}
{ \left< { \delta \phi   }^{ 2 } \right>  }_{ \rm ren }
\simeq \frac { 1 }{ 2{ \pi  }^{ 2 }C\left( \eta  \right)   } \int _{ 0 }^{ H }{  k^{2}{ \left| \delta { \chi  }_{ k } \right|  }^{ 2 }dk } 
+\frac { 1 }{ 2{ \pi  }^{ 2 }C\left( \eta  \right)   } \int _{ H }^{ \sqrt { 2 } /\left| \eta  \right|  }{  k^{2}{ \left| \delta { \chi  }_{ k } \right|  }^{ 2 } dk } \label{eq:fkhdg}, 
\end{eqnarray}
as well as the massless minimally coupled case.
At the late-time limit ($\eta\simeq0$ or $Ht\gg1$), the first integral converges to zero
(see Ref.\cite{Haro:2010zz,Haro:2010mx} for the details) and obtain the following approximation 
\begin{eqnarray}
{ \left< { \delta \phi   }^{ 2 } \right>  }_{ \rm ren }
&\simeq&\frac { 1 }{ 2{ \pi  }^{ 2 }C\left( \eta  \right)   } \int _{ H }^{ \sqrt { 2 } /\left| \eta  \right|  }
{  k^{2}{ \left| \delta { \chi  }_{ k } \right|  }^{ 2 } dk },\nonumber \\
&\simeq& \frac { 3{ H }^{ 4 } }{ 8{ \pi  }^{ 2 }{ m }^{ 2 } } \left[ 1-{ e }^{ -2{ m }^{ 2 }t/3H } \right]  \label{eq:fkhfgedg}.
\end{eqnarray}
which coincides with Eq.~(\ref{eq:gmkmkmg}) by using the physical cut-off.

Then, we briefly discuss the renormalized vacuum fluctuations ${ \left< { \delta \phi   }^{ 2 } \right>  }_{ \rm ren }$ 
for $m\gg H$ by using the adiabatic regularization. 
In this case, the power spectrum on super-horizon scales is approximately given by~\cite{Riotto:2002yw}
\begin{equation}
 { \Delta   }_{ \delta \phi  }^{2}\left( k \right)  = { \left( \frac { H }{ 2\pi  }  \right)  }^{ 2 }
\left( \frac { H }{ m }  \right) { \left( \frac { k }{ aH }  \right)  }^{ 3}\label{eq:fkh;fkhdg}.
\end{equation}
For the very massive case $m\gg H$, the amplitude of the power spectrum is suppressed 
and the spectrum on long wave modes rapidly drops down. Therefore, the massive field vacuum fluctuations 
are more inhomogeneous fluctuations and then, break the scale invariance of the spectrum of perturbations. 
In this case, we must pay attention to the UV cut-off.
For the very massive case $m\gg H$, the adiabatic conditions $\Omega^{2}_{k}>0$ satisfy for all $k$ modes and we can estimate 
the renormalized vacuum fluctuations by eliminating the lowest-order adiabatic (WKB) approximation 
\begin{eqnarray}
{ \left< { \delta \phi   }^{ 2 } \right>  }_{ \rm ren }
&\simeq&{ \left< {  \delta \phi   }^{ 2 } \right>  }_{W_{k}}-{ \left< {  \delta \phi   }^{ 2 } \right>  }_{\Omega_{k}},\\
&=&\frac { 1 }{ 4{ \pi  }^{ 2 }C\left( \eta  \right)  }  
\Biggl[ \int _{ 0 }^{ \Lambda  }{{ \frac { { k }^{ 2 } }{ { W }_{ k } } dk }}-\int _{ 0 }^{ \Lambda  }
{ \frac { { k }^{ 2 } }{ { \Omega }_{ k } } dk } \Biggr] \label{eq:ofhgjgjrg}.
 \end{eqnarray}
By using Eq.~(\ref{eq:oekgkf}), we can simply estimate the dominated terms as follows:
\begin{eqnarray}
&&\lim _{ \Lambda \rightarrow \infty  }\frac { 1 }{ 4{ \pi  }^{ 2 }C\left( \eta  \right)  }  
\Biggl[ \frac { { m }^{ 2 }C'' }{ 8 } \int _{ 0 }^{ \Lambda  }{ \frac { { k }^{ 2 } }{ { \omega  }_{ k }^{ 5 } } dk
-\frac { { 5m }^{ 4 }{ \left( C' \right)  }^{ 2 } }{ 32 } \int _{ 0 }^{ \Lambda  }{ \frac { { k }^{ 2 } }{ { \omega  }_{ k }^{ 7 } } dk }  }\Biggr]
\nonumber \\
&&=\lim _{ \Lambda \rightarrow \infty  }\frac { 1 }{ 4{ \pi  }^{ 2 }C\left( \eta  \right)  }  \Biggl[ \frac { { m }^{ 2 }C'' }{ 8 } \frac { { \Lambda  }^{ 3 } }{ 3{ m }^{ 2 }C{ \left( { \Lambda  }^{ 2 }+{ m }^{ 2 }C \right)  }^{ 3/2 } } 
-\frac { 5{ m }^{ 4 }{ \left( C' \right)  }^{ 2 } }{ 32 } \frac { { { 5m }^{ 2 }C\Lambda  }^{ 3 }+2{ \Lambda  }^{ 5 } }{ 15{ m }^{ 4 }{ C }^{ 2 }{ \left( { \Lambda  }^{ 2 }+{ m }^{ 2 }C \right)  }^{ 5/2 } }   \Biggr], \nonumber \\
&&=-\frac { 1 }{ 96{ \pi  }^{ 2 }C\left( \eta  \right)  } 
\left[ \frac { 1 }{ 2 } { \left( \frac { C' }{ C }  \right)  }^{ 2 }-\frac { C'' }{ C }  \right]=\frac { 1 }{ 48{ \pi  }^{ 2 } } \frac { a'' }{ { a }^{ 3 } }
=\frac{R}{288\pi^2}\label{eq:fhkgkgedg}.
\end{eqnarray}

\section{Renormalized vacuum fluctuations via point-splitting regularization}
\label{sec:point}
The point-splitting regularization is the method of regularizing divergences as the point separation 
in the two-point function, and has been studied in detail in Ref.\cite{Birrell:1982ix,Bunch:1978yq,Vilenkin:1982wt}.
In this section, we review the renormalized vacuum fluctuations by using the point-splitting regularization,
and compare the results in the previous section.
The regularized vacuum expectation values are expressed as~\cite{Vilenkin:1982wt}
\begin{eqnarray}
\begin{split}
 \left< {\delta \phi  }^{ 2 } \right>_{\rm reg}  =& -16{ \pi  }^{ 2 }{ \epsilon  }^{ 2 }+\frac { R }{ 576{ \pi  }^{ 2 } } +\frac { 1 }{ 16{ \pi  }^{ 2 } } \left[ { m }^{ 2 }+\left( \xi -\frac { 1 }{ 6 }  \right) R \right] \\ & \left[ \ln { \left( \frac { { \epsilon  }^{ 2 }{ \mu  }^{ 2 } }{ 12 }  \right)  } +\ln { \left( \frac { R }{ { \mu  }^{ 2 } }  \right) +2\gamma -1+\psi \left( \frac { 3 }{ 2 } +\nu  \right)  } +\psi \left( \frac { 3 }{ 2 } -\nu  \right)  \right] .
\end{split}\label{eq:hfjhkdg}
\end{eqnarray}
where we take Bunch-Davies vacuum state, $\epsilon $ is the regularization parameter which corresponds with the point separation, 
$\mu$ is the renormalization scale, $\gamma $ is Euler's constant and $\psi  \left( z \right)=\Gamma' \left( z \right)/\ \Gamma \left( z \right)$ is the digamma function. By using $m^{2}$ and $\xi$ renormalization, we have the following expression
\begin{eqnarray}
\begin{split}
\left< { \delta \phi    }^{ 2 } \right>_{\rm ren} & = \frac { 1 }{ 16{ \pi  }^{ 2 } }\biggr\{-{ m }^{ 2 }\ln { \left( \frac { 12{ m }^{ 2 } }{ { \mu  }^{ 2 } }  \right)}\\ &  +\left[ { m }^{ 2 }+\left( \xi -\frac { 1 }{ 6 }  \right) R \right] \left[ \ln { \left( \frac { R }{ { \mu  }^{ 2 } }  \right) +\psi \left( \frac { 3 }{ 2 } +\nu  \right)  } +\psi \left( \frac { 3 }{ 2 } -\nu  \right)  \right] \biggr\}.
\end{split}\label{eq:hitihiegedg}
\end{eqnarray}
where the additive constant $\psi \left( \frac { 3 }{ 2 } \pm \nu  \right)$ has been chosen so that $\left< {  \delta \phi   }^{ 2 } \right>_{\rm ren}   =0$ at the radiation-dominated Universe $R=0$.
In the massive field theory, we can remove simply the renormalization scale $\mu$ to set $\mu^{2}=12m^{2}$
\begin{eqnarray}
\left< {  \delta \phi  }^{ 2 } \right>_{\rm ren}  = \frac { 1 }{ 16{ \pi  }^{ 2 } }\left[ { m }^{ 2 }+\left( \xi -\frac { 1 }{ 6 }  \right) R \right] \left[ \ln { \left( \frac { R }{  { 12m }^{ 2 } }  \right) +\psi \left( \frac { 3 }{ 2 } +\nu  \right)  } +\psi \left( \frac { 3 }{ 2 } -\nu  \right)  \right]\label{eq:klkrkkdg}
\end{eqnarray}
In the massless and nearly conformal coupling case $m=0$ and $\xi \simeq1/6$, 
we cannot remove the renormalization scale $\mu$ and
$\psi \left( \frac { 3 }{ 2 } \pm \nu  \right)$ may be absorbed by the non-minimal coupling renormalization
\begin{equation}
\left< {  \delta \phi   }^{ 2 } \right>_{\rm ren}  =\frac { \xi -1/6 }{ 16{ \pi  }^{ 2 } } R\ln { \left( \frac { R }{ { \mu  }^{ 2 } }  \right)}\label{eq:klkorkhkdg}.
\end{equation}
In the massless conformal coupling case $m=0$ and $\xi=1/6$, we can simply obtain the following expression~\footnote{ This is equal to the thermal fluctuations $\left< { \delta \phi    }^{ 2 } \right>_{\rm ren} =T^{2}/12$ with the Gibbons-Hawking temperature $T=H/2\pi$ experienced by a point observer inside the de Sitter horizon (see, e.g. Ref.\cite{Acquaviva:2014xoa,Obadia:2008rt}).}
\begin{equation}
\left< { \delta \phi    }^{ 2 } \right>_{\rm ren}  =\frac { R }{ 576{ \pi  }^{ 2 } }=\frac { H^{2} }{ 48{ \pi  }^{ 2 }}\label{eq:hgl;l;fldg}.
\end{equation}
In the minimal coupling case $\xi=0$ and we take massless limit $m\rightarrow 0$
\begin{equation}
\left< {  \delta \phi   }^{ 2 } \right>_{\rm ren}  \rightarrow\frac { R^{2} }{ 384{ \pi  }^{ 2 }m^{2} }=\frac { 3H^{4} }{ 8{ \pi  }^{ 2 }m^{2} }\label{eq:hl;lhf:kdg}.
\end{equation}
which corresponds with Eq.~(\ref{eq:fkhfgedg}).
Then, the digamma function $\psi  \left( z \right)$ for $z\gg1$ can be approximated  as~\cite{Bunch:1978yq}
\begin{equation}
{\rm Re} \ {\psi  \left( \frac{3}{2}+i z \right)}= \log { z } +\frac { 11 }{ 24{ z }^{ 2 } } -\frac { 127 }{ 960{ z }^{ 4 } } +\cdots \label{eq:flklkdg}.\end{equation}
In the massive case $m\gg H$, 
\begin{eqnarray}
&&\ln { \left( \frac { { H }^{ 2 } }{ { m }^{ 2 } }  \right)  } +\psi \left( \frac { 3 }{ 2 } +\nu  \right) +\psi \left( \frac { 3 }{ 2 } -\nu  \right)\nonumber \\
\approx && \ln { \left( \frac { { H }^{ 2 } }{ { m }^{ 2 } }  \right)  } +\psi \left( \frac { 3 }{ 2 } +i\frac { m }{ H }  \right) +\psi \left( \frac { 3 }{ 2 } -i\frac { m }{ H }  \right),\\
 \approx &&\frac { 11 }{ 12 } \frac { { H }^{ 2 } }{ { m }^{ 2 } } -\frac { 127 }{ 480 } \frac { { H }^{ 4 } }{ { m }^{ 4 } } +\cdots \label{eq:dlkglkdg}.\end{eqnarray}
Therefore, the renormalized vacuum fluctuations of the massive Higgs field for $m\gg H$ is given as follows:
\begin{eqnarray}
\left< {  \delta \phi  }^{ 2 } \right>_{\rm ren}  &=& \frac { 1 }{ 16{ \pi  }^{ 2 } }\left[ { m }^{ 2 }+\left( \xi -\frac { 1 }{ 6 }  \right) R \right] \left(\frac { 11 }{ 12 } \frac { { H }^{ 2 } }{ { m }^{ 2 } } -\frac { 127 }{ 480 } \frac { { H }^{ 4 } }{ { m }^{ 4 } } +\cdots  \right), \\
&\simeq& O\left( H^{2} \right) \label{eq:hldfkldg},
\end{eqnarray}
which is consistent with Eq.~(\ref{klklksssg}).
Therefore, the renormalized vacuum fluctuations $\left< { \delta \phi    }^{ 2 } \right>_{\rm ren}$ 
via the point-splitting regularization is equivalent to the results of Eq.~(\ref{eq:fhkgkgedg}) via the adiabatic regularization.
Finally, we summarize the renormalized vacuum fluctuations ${ \left< { \delta \phi   }^{ 2 } \right>  }_{ \rm ren }$ 
via the adiabatic regularization and the point-splitting regularization as follows:
\begin{equation}
\left< {  \delta \phi  }^{ 2 } \right>_{\rm ren}\simeq\begin{cases} { H }^{ 3 }t /4{ \pi  }^{ 2 } ,\quad\quad\quad  \left( m=0 \right) 
\\  3{ H }^{ 4} / 8{ \pi  }^{ 2 }m^{2},\quad\ \left( m\ll H \right) \\  H^2/24\pi^{2}. \quad\quad\ \ \left( m\gtrsim  H \right) \end{cases} \label{klklksssg}
\end{equation}

\section{Electroweak vacuum instability from dynamical behavior of homogeneous Higgs field 
and renormalized Higgs vacuum fluctuations}
\label{sec:electroweak}
In this section, we apply the results of Section~\ref{sec:effective}, Section~\ref{sec:adiabatic}
and Section~\ref{sec:point} to the investigation of the electroweak 
vacuum instability during inflation (de Sitter space) or after inflation. 
The vacuum instability of the electroweak false vacuum on the dynamical background
is determined by the behavior of the homogeneous Higgs field $\phi$ and the inhomogeneous Higgs field fluctuations,
i.e., the renormalized vacuum fluctuations $\left< {  \delta \phi  }^{ 2 } \right>_{\rm ren}$.
The behavior of the global and homogeneous Higgs field $\phi$ is governed by 
the effective Klein-Gordon equation
\begin{eqnarray}
\Box \phi + V'_{\rm eff}\left( \phi  \right) =0\label{eq:gkfkjdkdg}.
\end{eqnarray}
We can rewrite the effective Klein-Gordon equation as the following
\begin{eqnarray}
\ddot { \phi }\left(t\right)  +3H\dot { \phi  }\left(t\right)  +V'_{\rm eff}\left( \phi \left(t\right)  \right) =0. \label{eq:kdljkdddljlkdj}
\end{eqnarray}
If we approximate the effective potential as $V'_{\rm eff}\left( \phi  \right) =-m^{2}_{\rm eff}\phi$, 
the behavior of the coherent Higgs field $\phi\left(t\right) $ is described as follows:
\begin{eqnarray}
\phi\left(t\right)  \propto { e }^{ \frac { 1 }{ 2 } \left( -3H+\sqrt { 9{ H }^{ 2 }+4m^{2}_{\rm eff} }  \right) t }
\simeq\begin{cases} { e }^{ m_{\rm eff}t }  \ \ \quad\quad \left( m_{\rm eff}\gg  H \right),
\\  { e }^{ m^{2}_{\rm eff}t/3H }  \quad \ \left( m_{\rm eff}\lesssim  H \right).\end{cases} \label{eq:kdljklkdj}
\end{eqnarray}

The one-loop standard model Higgs effective potential $V_{\rm eff}\left( \phi  \right)$ of curved space-time
~\cite{Herranen:2014cua} in the 't Hooft-Landau gauge and the $\overline {\rm  MS } $ scheme, is given as follows:
\begin{eqnarray}
V_{\rm eff}\left( \phi  \right) &&=\frac{1}{2}m^{2}(\mu)\phi^{2}+\frac{1}{2}\xi (\mu)R\phi^{2}
+\frac{\lambda(\mu)}{4}\phi^{4}\\ &&+\sum _{ i=1 }^{ 9 }{ \frac { { n }_{ i } }{ 64{ \pi  }^{ 2 } } { M }_{ i }^{ 4 }\left( \phi  \right) \left[ \log{ \frac { { M }_{ i }^{ 2 }\left( \phi  \right)  }{ { \mu  }^{ 2 } }  } -{ C }_{ i } \right]  } \nonumber \label{eq:fhlgkdg},
\end{eqnarray}
where
\begin{equation}
{ M }_{ i }^{ 2 }\left( \phi  \right) ={ \kappa  }_{ i }{ \phi  }^{ 2 }+{ \kappa }'_{ i }+\theta_{i} R.
\end{equation}
The coefficients $n_{i}$, ${ \kappa  }_{ i }$, ${ \kappa }'_{ i }$ and $\theta_{i}$ 
are given by Table I of Ref.\cite{Herranen:2014cua}. 
Furthermore, the $\beta$-function for the non-minimal coupling $\xi\left( \mu  \right)$  
in the standard model ignoring gravity, is given by
\begin{equation}
\beta_{\xi }=\frac { 1  }{ { \left( 4\pi  \right)  }^{ 2 } }\left( \xi -1/6 \right) 
\left( 6\lambda +3{ y }_{ t }^{ 2 }-\frac { 3 }{ 4} { g' }^{ 2 }-\frac { 9 }{ 4 } { g }^{ 2 } \right)  \label{eq:sdjldkdg}.
\end{equation}
The running of the non-minimal coupling $\xi \left( \mu  \right) $ can be obtained by integrating $\beta_{\xi }$
\begin{equation}
\xi \left( \mu  \right) =\frac { 1 }{ 6 } +\left( { \xi  }_{ \rm EW }-\frac { 1 }{ 6 }  \right) F\left( \mu  \right)  \label{eq:gdjglg},
\end{equation}
where $F\left( \mu  \right)$ depend on the renormalization scale $\mu$.
If we have the nearly minimal coupling ${ \xi  }_{ \rm EW }\lesssim O\left(10^{-2}\right)$ at the electroweak scale, 
the running non-minimal coupling $\xi \left( \mu  \right) $ becomes negative at some scale~\cite{Herranen:2014cua}.
On the other hand, we can take the initial condition of the running non-minimal coupling $\xi \left( \mu  \right) $ 
at the Planck scale~\cite{Espinosa:2015qea}. If the Hubble scale is larger than the instability scale
\footnote{ We define the instability scale $\Lambda_{I}$ as the derivative of
the standard model Higgs effective potential in Minkowski space-time becomes negative at the scale
and the current experiments of the Higgs boson mass,
$m_{h}=125.09\ \pm \  0.21\ ({\rm stat})\ \pm \  0.11\ ({\rm syst})\ {\rm GeV}$
~\cite{Aad:2015zhl,Aad:2013wqa,Chatrchyan:2013mxa,Giardino:2013bma}
and top quark mass, $m_{t}=173.34\pm 0.27\ ({\rm stat})\pm 0.71\ ({\rm syst})\ {\rm GeV}$~\cite{2014arXiv1403.4427A} 
suggest the instability scale $\Lambda_{I} = 10^{10} \sim 10^{11} \ {\rm GeV}$~\cite{Buttazzo:2013uya}.}
i.e., $H>\Lambda_{I}$ and $\xi \left(  H \right)<0 $, the Higgs effective potential 
in de Sitter space becomes negative, i.e., $V_{\rm eff}'\left( \phi  \right)\lesssim 0$, and the homogeneous 
Higgs field $\phi$ on the entire Universe rolls down to the negative-energy Planck-scale true vacuum.

However, as mentioned in Section~\ref{sec:effective}, we must shift the Higgs field 
$\phi^{2}\rightarrow \phi^{2}+\left< {  \delta \phi  }^{ 2 } \right>_{\rm ren} $
in order to include the backreaction term via the renormalized vacuum fluctuations and 
the one-loop standard model Higgs effective potential in curved space-time is modified as follows: 
\begin{eqnarray}
V_{\rm eff}\left( \phi  \right) &&=\frac{1}{2}m^{2}(\mu)\phi^{2}+\frac{1}{2}\xi (\mu)R\phi^{2}
+\frac{1}{2}\lambda(\mu)\left< {\delta \phi  }^{ 2 } \right>_{\rm ren}\phi^{2}+\frac{\lambda(\mu)}{4}\phi^{4} \label{eq:gkljkkdg}
\\ &&+\sum _{ i=1 }^{ 9 }{ \frac { { n }_{ i } }{ 64{ \pi  }^{ 2 } } { M }_{ i }^{ 4 }\left( \phi  \right) \left[ \log{ \frac { { M }_{ i }^{ 2 }\left( \phi  \right)  }{ { \mu  }^{ 2 } }  } -{ C }_{ i } \right]  } \nonumber.
\end{eqnarray}
with
\begin{equation}
{ M }_{ i }^{ 2 }\left( \phi  \right) ={ \kappa  }_{ i }{ \phi  }^{ 2 }+{ \kappa  }_{ i }
\left< {\delta \phi  }^{ 2 } \right>_{\rm ren}+{ \kappa }'_{ i }+\theta_{i} R\label{eq:gkldfdfkdg}.
\end{equation}
Here, we must choose the appropriate scale $\mu$ in order to suppress the higher order corrections
of $\log{  { { M }_{ i }^{ 2 }\left( \phi  \right)  }/{ { \mu  }^{ 2 } }  }$.
In the case of flat spacetime $R=0$, it is known that $\mu^{2} =\phi^{2}$ is a good choice to suppress the 
high order log-corrections. In de Sitter space, however, we must choose $\mu^{2} =\phi^{2}+R+\left< {\delta \phi  }^{ 2 } \right>_{\rm ren}=\phi^{2}+12H^{2}+\left< {\delta \phi  }^{ 2 } \right>_{\rm ren}$ to suppress the log-corrections. 
Therefore, if we have $\mu\simeq{ \left(  12H^{2}+\left< {\delta \phi  }^{ 2 } \right>_{\rm ren}\right)  }^{ 1/2 }>\Lambda_{I}$,
the quartic term  $\lambda(\mu)\phi^{4}/4$ makes negative contribution to the effective potential.

The Higgs field phenomenologically acquires various effective masses during inflation or after inflation, e.g. 
the inflaton-Higgs coupling $\lambda_{\phi S}$ provides an extra contribution to the Higgs mass 
$m^{2}_{\rm eff}=\lambda_{\phi S}S^{2}$  where $S$ is the inflaton field. 
However, in this work, we restrict our attention to the simple case 
that the Higgs field only couples to the gravity via the non-minimal Higgs-gravity coupling $\xi(\mu)$ and 
we disregard other inflationary effective mass-terms. For convenience, we only consider the inflationary effective 
mass-term $m^{2}_{\rm eff}=\xi(\mu)R=12\xi(\mu)H^{2 }_{\rm inf}$ and use the results of Eq.~(\ref{klklksssg}), 
the renormalized vacuum field fluctuations are given by
\begin{equation}
\left< {  \delta \phi  }^{ 2 } \right>_{\rm ren}\simeq\begin{cases} { H }^{ 2}_{\rm inf} / 32{ \pi  }^{ 2 }\xi(\mu),
\quad\quad \left( \xi(\mu) \ll O\left(10^{-1}\right) \right) \\  
H^{2}_{\rm inf}/24\pi^{2}. \quad\quad\quad\quad\ \ \left( \xi(\mu)\gtrsim  O\left(10^{-1}\right) \right) \end{cases} \label{klksklkdj}
\end{equation}
Here, if we have $\mu\simeq{ \left(  12H^{2}_{\rm inf}+\left< {\delta \phi  }^{ 2 } \right>_{\rm ren}\right)  }^{ 1/2 }>\Lambda_{I}$
\footnote{ In the case ${ \left(  12H^{2}_{\rm inf}+\left< {\delta \phi  }^{ 2 } \right>_{\rm ren}\right)  }^{ 1/2 }<\Lambda_{I}$,
the quartic term  $\lambda(\mu)\phi^{4}/4$ makes positive contribution to the effective potential 
unless $\phi>\Lambda_{I}$. Therefore, the homogeneous Higgs field $\phi$ cannot classically go out to the Planck-scale vacuum state. However, it is possible to generate AdS domains or bubbles via the inhomogeneous Higgs field fluctuations shown in (\ref{eq:hswedg}).},
we can infer the sign of $V_{\rm eff}\left( \phi  \right)$ by using the relations 
$\xi (\mu)R<\left| {\lambda(\mu)}\left< {\delta \phi  }^{ 2 } \right>_{\rm ren} \right| $.
If we assume $\xi (\mu)R=\xi (\mu)12H^{2}_{\rm inf}$, ${\lambda(\mu)}\left< {\delta \phi  }^{ 2 } \right>_{\rm ren}
\simeq{\lambda(\mu)}{ H }^{ 2}_{\rm inf} / 32{ \pi  }^{ 2 }\xi(\mu)$ and $\lambda(\mu)\simeq -0.01$, 
we obtain the constraint on the non-minimal coupling
${ \xi (\mu) }\lesssim O\left(10^{-3}\right)$ where $H_{\rm inf}>\sqrt { 32{ \pi  }^{ 2 }\xi \left( \mu  \right)  }\ \Lambda_{I} $. 
In this case, the homogeneous Higgs field $\phi$ goes out to
the negative Planck-energy vacuum state, and therefore, the excursion of the homogeneous Higgs field $\phi$ 
to the Planck-scale true vacuum can terminate inflation and cause an immediate collapse of the Universe.

On the other hand, the inhomogeneous Higgs field fluctuations $\left< {  \delta \phi  }^{ 2 } \right>_{\rm ren}$
can cause the vacuum transition of the Universe~\cite{Espinosa:2007qp,Fairbairn:2014zia,Lebedev:2012sy,Kobakhidze:2013tn,Enqvist:2013kaa,Herranen:2014cua,Kobakhidze:2014xda,Kohri:2016wof,Kamada:2014ufa,Enqvist:2014bua,Hook:2014uia,Kearney:2015vba,Espinosa:2015qea}. If the inhomogeneous and local Higgs field gets over the 
hill of the effective potential, the local Higgs fields classically roll down into the negative Planck-energy true vacuum 
and catastrophic Anti-de Sitter (AdS) domains are formed.
Note that not all AdS domains formed during inflation threaten the existence of the Universe~\cite{Hook:2014uia, Espinosa:2015qea}, which highly depends on the evolution of the AdS domains at the end of inflation (see Ref.\cite{Espinosa:2015qea} for the details). The AdS domains can either shrink or expand, eating other regions of the electroweak vacuum. Although the high-scale inflation can generate more expanding AdS domains than shrinking domains during inflation, 
such domains never overcome the inflationary expansion of the Universe, i.e.,
one AdS domain cannot terminate the inflation of the Universe~\footnote{
The expansion of AdS domains or bubbles never takes over the expansion
of the inflationary dS space~\cite{Espinosa:2015qea}, and therefore, it is
impossible that one AdS domain terminates the inflation of the Universe. 
However, if the non-inflating domains or the AdS domains dominates 
all the space of the Universe~\cite{Sekino:2010vc}, 
the inflating space would crack, and the inflation of all the space of the Universe finally comes to an end.}.
However, after inflation, some AdS domains expand and consume the entire Universe. Therefore, the existence of AdS domains on our Universe is catastrophic, and so we focus on the conditions not to be generated during or after inflation.

We assume that the probability distribution function of the Higgs field fluctuations is Gaussian, i.e.
\begin{equation}
P\left( \phi, \left< {  \delta \phi  }^{ 2 } \right>_{\rm ren} \right) 
= \frac { 1 }{ \sqrt {2{ \pi  }\left< {  \delta \phi  }^{ 2 } \right>_{\rm ren} } } 
\exp \left( -\frac {{  \phi  }^{ 2 } }{ 2\left< {  \delta \phi  }^{ 2 } \right>_{\rm ren} }  \right)\label{eq:hqqqqdg}.
\end{equation}
By using Eq.~(\ref{eq:hqqqqdg}), 
the probability that the electroweak vacuum survives can be obtained as
\begin{eqnarray}
{ P }\left(  \phi<{ \phi }_{ \rm max }\right) &\equiv& \int _{ -{ \phi  }_{\rm  max } }^{ { \phi }_{ \rm max } }
{ P\left( \phi, \left< {  \delta \phi  }^{ 2 } \right>_{\rm ren} \right) d\phi }, \\ &=& {\rm erf}\left( \frac{{ \phi }_{ \rm max }  }
{ \sqrt { 2\left< {  \delta \phi  }^{ 2 } \right>_{\rm ren}  }}  \right)\label{eq:hswegg}.
\end{eqnarray}
where ${ \phi }_{ \rm max }$ defined as the Higgs effective potential $V_{\rm eff}\left( \phi  \right)$ 
given by Eq.~(\ref{eq:gkljkkdg}) takes its maximal value
\footnote{ The Higgs effective potential with the large effective mass-term can be approximated by
\[ V_{\rm eff }\left(  \phi \right) \simeq \frac { 1 }{ 2 }m^{2}_{\rm eff}{ \phi }^{ 2 }\left( 1-\frac { 1 }{ 2 } { \left( \frac {  \phi }{ { \phi}_{\rm  max } }  \right)  }^{ 2 } \right), \] 
where ${  \phi }_{\rm  max }$ is given by
\[{\phi }_{ \rm max }=\sqrt { -\frac { m^{2}_{\rm eff}}{ { \lambda }\left( \mu  \right)} }. \]
Our approximation $\phi_{\rm max}\simeq 10\ m_{\rm eff}$ is numerically valid for the one-loop effective potential.}.

On the other hand, the probability that the inhomogeneous Higgs field falls into the true vacuum can be expressed as
\begin{eqnarray}
{ P }\left(  \phi>{ \phi }_{ \rm max }\right)&=&1- {\rm erf}\left( \frac{{ \phi }_{ \rm max }  }
{ \sqrt { 2\left< {  \delta \phi  }^{ 2 } \right>_{\rm ren}  }}  \right), \\ &\simeq&\frac {\sqrt{2\left< {  \delta \phi  }^{ 2 } \right>_{\rm ren}  }}{\pi{ \phi }_{ \rm max }}e^{-\frac {{ \phi }_{ \rm max }^{2}}{2\left< {  \delta \phi  }^{ 2 } \right>_{\rm ren}}}\label{eq:afklsjiijgg}.
\end{eqnarray}
The vacuum decay probability is given by
\begin{equation}
{ e }^{ 3{ N }_{ \rm hor } }{ P }\left(  \phi>{ \phi }_{ \rm max }\right)<1,\label{eq:aaawegg}
\end{equation}
where ${ e }^{ 3{ N }_{ \rm hor } }$ corresponds to the physical volume of our universe at the
end of the inflation, and we take the e-folding number $N_{\rm hor }\simeq N_{\rm CMB }\simeq60$. Imposing Eq.~(\ref{eq:afklsjiijgg})
on the condition shown in (\ref{eq:aaawegg}), we obtain the relation
\begin{equation}
\frac{\left< {  \delta \phi  }^{ 2 } \right>_{\rm ren}}{{ \phi }_{ \rm max }^{2}}<\frac{1}{6N_{\rm hor} }\label{eq:hswedg}.
\end{equation}

Now, we consider the condition shown in (\ref{eq:hswedg}) by using the Higgs effective potential in de-Sitter space-time
given by Eq.~(\ref{eq:gkljkkdg}) and the Higgs field vacuum fluctuations by Eq.~(\ref{klklksssg}).
In the same way as our previous work~\cite{Kohri:2016wof}, 
we can numerically obtain the constraint of the non-minimal coupling ${ \xi (\mu) }\lesssim O\left(10^{-2}\right)$ 
where $H_{\rm inf}>\sqrt { 32{ \pi  }^{ 2 }\xi \left( \mu  \right)  }\ \Lambda_{I} $.
Therefore, the catastrophic Anti-de Sitter (AdS) domains or bubbles from the high-scale inflation can be avoided 
if the relatively large non-minimal Higgs-gravity coupling (or e.g. inflaton-Higgs coupling $\lambda_{\phi S}$) is introduced.
Here, we summarize the conclusions obtained in this section as follows:
\begin{itemize}
 \item In ${ \xi (\mu) }\lesssim O\left(10^{-3}\right)$ and $H_{\rm inf}>\sqrt { 32{ \pi  }^{ 2 }\xi \left( \mu  \right)  }\ \Lambda_{I} $, 
 the Higgs effective potential during inflation is destabilized by the backreaction 
 term ${\lambda(\mu)}\left< {\delta \phi  }^{ 2 } \right>_{\rm ren}$, which overcome the stabilization term $\xi (\mu)R$.
 In this case, the effective potential becomes negative, i.e. $V_{\rm eff}'\left( \phi  \right)\lesssim 0$, the excursion of the homogeneous Higgs field $\phi$ to the negative Planck-energy vacuum state terminates the inflation of the Universe 
 and cause a catastrophic collapse.
\item In $O\left(10^{-3}\right)\lesssim{ \xi (\mu) }\lesssim O\left(10^{-2}\right)$ and 
$H_{\rm inf}>\sqrt { 32{ \pi  }^{ 2 }\xi \left( \mu  \right)  }\ \Lambda_{I} $, 
The inflationary effective mass $\xi (\mu)R=12\xi (\mu)H^{2}_{\rm inf}$  can raise and stabilize 
the Higgs effective potential during inflation.
Thus, the dangerous motion of the homogeneous Higgs field $\phi$ cannot occur. However, the inhomogeneous Higgs field fluctuations $\left< {  \delta \phi  }^{ 2 } \right>_{\rm ren}$ generate the catastrophic Anti-de Sitter (AdS) domains or bubbles, 
which finally cause the vacuum transition of the Universe.
\item In ${ \xi (\mu) }\gtrsim  O\left(10^{-2}\right)$ or $H_{\rm inf}<\sqrt { 32{ \pi  }^{ 2 }\xi \left( \mu  \right)  }\ \Lambda_{I} $, 
the Higgs effective potential stabilizes and the catastrophic Anti-de Sitter (AdS) domains or bubbles are not formed during inflation.
\end{itemize}

However, after inflation, the effective mass-terms $\xi (\mu)R$
via the non-minimal Higgs-gravity coupling drops rapidly and sometimes become negative.
Therefore, the effect of the stabilization via $\xi (\mu)R$ disappears and the Higgs effective potential 
becomes rather unstable due to the terms of ${\lambda(\mu)}\left< {\delta \phi  }^{ 2 } \right>_{\rm ren}$
or $\xi (\mu)R$ with $\mu\simeq{ \left(  R+\left< {\delta \phi  }^{ 2 } \right>_{\rm ren}\right)  }^{ 1/2 }>\Lambda_{I} $.
Furthermore, the non-minimal Higgs-gravity coupling can generate the large Higgs field vacuum fluctuations 
via tachyonic resonance~\cite{Kohri:2016wof,Herranen:2015ima, Ema:2016kpf}, thus, the Higgs effective potential is destabilized, or the catastrophic Anti-de Sitter (AdS) domains or bubbles are formed during subsequent preheating stage. In the rest of this section, we briefly discuss the instability of the electroweak false vacuum after inflation.

Just after inflation, the inflaton field $S$ begins coherently oscillating near the minimum of the inflaton potential $V_{\rm inf}\left( S \right)$ and produces extremely a huge amount of massive bosons via the parametric or the tachyonic resonance. 
This temporal non-thermal stage is called preheating~\cite{Kofman:1997yn}, 
and is essentially different from the subsequent stages of the reheating and the thermalization.
For simplicity, we approximate the inflaton potential as the quadratic form 
\begin{equation}
V_{\rm inf}\left( S \right) =\frac { 1 }{ 2 } { m }_{S}^{ 2 }{ S }^{ 2 }.\label{eq:assjfkjwegg}
\end{equation}
In this case, the inflaton field $S$ classically oscillates as 
\begin{equation}
S\left( t \right) =\Phi \sin { \left( { m }_{ S }t\right) }, \quad \Phi 
=\sqrt { \frac { 8 }{ 3 }  } \frac { { M }_{ \rm pl } }{ { m }_{ S }t }, \label{eq:aslsjdksg}
\end{equation}
where the reduced Planck mass is ${ M }_{ \rm pl }  = 2.4\times 10^{18}\ { \rm GeV }$.
If the inflaton field $S$ dominates the energy density and the pressure of the Universe, i.e., 
during inflation or preheating stage, the scalar curvature $R(t)$ is written by
\begin{eqnarray}
R\left( t \right) &=&\frac { 1 }{ { M }_{\rm pl }^{ 2 } } \left[ 4V_{\rm inf}\left( S \right) -{ \dot { S }  }^{ 2 } \right],\\
&\simeq&\frac { { { m }_{ S }^{ 2 } }{ \Phi }^{ 2 } }{ { M }_{\rm  pl }^{ 2 } } \left( 3\sin^{2}\left( { { m }_{ S }t }\right) -1 \right). \label{eq:asjsljfgg}
\end{eqnarray}
When the inflaton field $S$ oscillates as Eq.~(\ref{eq:aslsjdksg}), the effective mass $\xi (\mu)R$ drastically changes
between positive and negative values. Therefore, the Higgs field vacuum fluctuations grows extremely rapidly
via the tachyonic resonance, which is called geometric preheating~\cite{Bassett:1997az,Tsujikawa:1999jh}.
The general equation for $k$ modes of the Higgs field during preheating is given as follows:
\begin{equation}
\begin{split}
\frac {d^{2}\left(a^{3/2}\delta { \phi  }_{ k } \right)}{  dt^{2}}+\left(\frac{k^{2} }{a^{2}}+V'_{\rm eff}\left(\phi  \right) +
\frac{1}{M_{\rm pl}^{2}} \left(\frac{3}{8}-\xi \right)\dot { S }\right. \\ \left.-\frac{1}{M_{\rm pl}^{2}} \left(\frac{3}{4}-4\xi \right)V\left( S \right) \right)
\left(a^{3/2}\delta { \phi  }_{ k }  \right)=0\label{eq:fhslsdfff}.
\end{split}
\end{equation}
Eq.~(\ref{eq:fhslsdfff}) can be reduced to the following Mathieu equation
\begin{equation}
\frac {d^{2}\left(a^{3/2}\delta  { \phi  }_{ k }  \right)}{  dz^{2}}+\left(A_{k}-2q\cos2z\right)
\left(a^{3/2}\delta  { \phi  }_{ k }  \right)=0,\label{eq:kmathieu}
\end{equation}
where we take $z=m_{S}t$ and $A_{k}$ and $q$ are given as 
\begin{eqnarray}
A_{k}&=&\frac{k^{2}}{a^{2}m_{S}^{2}}+\frac{V'_{\rm eff}\left(\phi \right)}{m_{S}^{2}}
+\frac{\Phi^{2}}{2 M_{\rm pl}^{2}}\xi, \\
q&=&\frac{3\Phi^{2}}{4M_{\rm pl}^{2}}\left(\xi-\frac{1}{4}\right).
\end{eqnarray}
The solutions of the Mathieu equation via the non-minimal coupling in Eq.~(\ref{eq:kmathieu}) 
show the tachyonic (broad) resonance when $q\gtrsim  1$ or the narrow resonance
when $q<1$. In the tachyonic resonance regime, where $q\gtrsim 1$, i.e. $ \Phi^{2}\xi \gtrsim M_{\rm pl}^{2}$, 
the tachyonic resonance extremely amplifies the Higgs vacuum fluctuations at the end of inflation as  $\left< {  \delta \phi  }^{ 2 } \right>_{\rm ren}\gg O\left(H^{2}\left(t \right)\right)$. In the context of preheating, $A_{k}$ and $q$ are $z$-dependent function due to the expansion of the Universe, making it very difficult to derive analytical estimation 
(see e.g. Ref.\cite{Ema:2016kpf}). 
If we take $m_{S}\simeq 7\times10^{-6}M_{\rm pl}^{2} $ assuming chaotic inflation with a quadratic potential, 
we can numerically obtain the condition of the tachyonic resonance as ${ \xi (\mu) }\gtrsim  O\left(10\right)$ 
(see e.g. Ref.\cite{Kohri:2016wof} for the details).
In the narrow resonance regime, where $q< 1$, i.e. $\Phi^{2}\xi < M_{\rm pl}^{2}$, 
the tachyonic resonance cannot occur, and therefore, the Higgs vacuum fluctuations after inflation 
decrease as $\left< {  \delta \phi  }^{ 2 } \right>_{\rm ren}\simeq O\left(H^{2}\left(t \right)\right)$ 
due to the expansion of the Universe. 
Here, we briefly summarize the results of the Higgs field vacuum fluctuations after inflation as follows:
\begin{eqnarray}
\begin{cases} \left< {  \delta \phi  }^{ 2 } \right>_{\rm ren}\gg O\left(H^{2}\left(t \right)\right),
\quad\quad \left(\Phi^{2}\xi \gtrsim M_{\rm pl}^{2} \right)\\ 
\left< {  \delta \phi  }^{ 2 } \right>_{\rm ren}\simeq O\left(H^{2}\left(t \right)\right).
\quad\quad\ \left(\Phi^{2}\xi < M_{\rm pl}^{2} \right) \end{cases} \label{eq:jkdljkdklkdj}
\end{eqnarray}
In the same way as the inflation stage, if we have 
$\mu\simeq{ \left(  R+\left< {\delta \phi  }^{ 2 } \right>_{\rm ren}\right)  }^{ 1/2 } >\Lambda_{I}$,
we shall infer the sign of $V_{\rm eff}\left( \phi  \right)$ by using the inequality $\xi (\mu)R\left(t\right)<\left| {\lambda(\mu)}\left< {\delta \phi  }^{ 2 } \right>_{\rm ren} \right| $, the scalar curvature $\left| R\left(t\right) \right| \simeq 3H^{2}\left(t\right)$ and the self-coupling $\lambda(\mu)\simeq -0.01$. If the tachyonic resonance happens, it is clear that the effective potential becomes negative, i.e. $V_{\rm eff}'\left( \phi  \right)\lesssim 0$, and then,
the homogeneous Higgs field $\phi$ goes out to the negative Planck-energy vacuum state.
On the other hand, in the narrow resonance, it cannot happen the same catastrophe, where 
$\xi (\mu)R\left(t\right)<\left| {\lambda(\mu)}\left< {\delta \phi  }^{ 2 } \right>_{\rm ren} \right| $, 
because the inhomogeneous Higgs field fluctuations after inflation are given as
$\left< {  \delta \phi  }^{ 2 } \right>_{\rm ren}\simeq O\left(H^{2}\left(t \right)\right)$ due to the expansion of the Universe, and 
the scalar curvature decreases as  $\left| R\left(t\right) \right| \simeq 3H^{2}\left(t\right) $.
Therefore, the narrow resonance does not destabilize the effective potential during preheating.

However, we recall that the scalar curvature $R \left( t \right) $ shown 
in (\ref{eq:asjsljfgg}) oscillates during each cycle $t\simeq 1/m_{S}$.
The stabilization of $\xi (\mu)R\left( t \right)$ to the coherent Higgs field generally does not work at the end of the inflation, 
because $\xi (\mu)R\left( t \right)$ changes sign during each oscillation cycle.
If the oscillation time-scale $t\simeq 1/m_{S}$ is relatively long, the curvature term $\xi (\mu)R\left( t \right)$ can
accelerate catastrophic motion of the coherent Higgs field $\phi\left( t \right)$ immediately after the end of the inflation.
Here, we briefly discuss the development of the coherent Higgs field $\phi\left( t \right)$ at the end of the inflation.
In one oscillation time-scale $t\simeq 1/m_{S}$, we can simply approximate the effective mass as 
$m^{2}_{\rm eff}\simeq \xi (\mu)R\left( t \right)\approx -3\xi (\mu)H^{2}_{\rm end}$.
By using Eq.~(\ref{eq:kdljklkdj}), 
the classical behavior of the coherent Higgs field $\phi\left( t \right)$ at the end of the inflation can be approximated as
\begin{eqnarray}
\phi\left( t \right)&\simeq& \phi_{\rm end } \cdot 
{ e }^{\left(3\xi (\mu)H^{2}_{\rm end} \right)t/3H_{\rm end}},\\
&\simeq&\phi_{\rm end } \cdot 
{ e }^{\left(3\xi (\mu)H^{2}_{\rm end}/3H_{\rm end}m_{S}  \right)},\\
&\simeq&\phi_{\rm end } \cdot 
{ e }^{\left(\xi (\mu)H_{\rm end}/m_{S} \right) },
\label{eq:jskjfdklkdj}
\end{eqnarray}
where the coherent Higgs field  $\phi_{\rm end }$ at the end of inflation is generally not zero, and 
corresponds to the Higgs field vacuum fluctuations 
$\left< {\delta \phi  }^{ 2 } \right>_{\rm ren} \simeq O\left( H^{2}_{\rm end}\right)$ at the end of inflation. 
Therefore, if we have $\phi\left( t \right)>\phi_{\rm max}$
\footnote{ By using $\phi_{\rm max}\simeq 10m_{\rm eff}$, we can simply approximate 
$\phi_{\rm max}\simeq 10\sqrt{\xi (\mu)\left| R\left(t\right) \right|}\simeq 10H_{\rm end}\sqrt{3\xi (\mu)}$
}, i.e. $H_{\rm end}/m_{S}\gtrsim (\log { 10\sqrt { 3\xi \left( \mu  \right)  } }) /\xi \left( \mu  \right)$,
the almost coherent Higgs fields $\phi\left( t \right)$
produced at the end of the inflation go out to the negative Planck-vacuum state 
and cause a catastrophic collapse of the Universe.

Furthermore, the inflation produces an enormous amount of causally disconnected horizon-size domains
and our observable Universe contains $e^{3N_{\rm hor}}$ of them.
Thus, we can consider one domain which has the large Higgs field fluctuations 
$6N_{\rm hor}\left< {\delta \phi  }^{ 2 } \right>_{\rm ren}$ by using Eq.~(\ref{eq:hswedg}).
The classical motion of the coherent Higgs field on such domain can be given as the following
\begin{eqnarray}
\phi\left( t \right)&\simeq& \sqrt {6N_{\rm hor}\left< {\delta \phi  }^{ 2 } \right>_{\rm ren}  }  \cdot 
{ e }^{\left(3\xi (\mu)H^{2}_{\rm end} \right)t/3H_{\rm end}},\\
&\simeq& 10H_{\rm end} \cdot { e }^{\left(\xi (\mu)H_{\rm end}/m_{S} \right) },
\label{eq:jsjdlssdklkdj}
\end{eqnarray}
where we take the e-folding number $N_{\rm hor}=60$. Therefore, if we have $\phi\left( t \right)>\phi_{\rm max}$
i.e., $H_{\rm end}/m_{S}\gtrsim ( \log { \sqrt { 3\xi \left( \mu  \right)  }  })/\xi \left( \mu  \right)$, 
the coherent Higgs field $\phi\left( t \right)$ on such domain goes out to the negative Planck-vacuum state and forms the
catastrophic AdS domains or bubbles, which finally cause the vacuum transition of the Universe.
That conclusion depends strongly on the non-minimal coupling $\xi(\mu)$,
the oscillation time-scale $t\simeq 1/m_{S}$ and the Hubble scale $H_{\rm end}$ at the end of the inflation.
Thus, the large non-minimal Higgs-gravity coupling $\xi(\mu)$ can destabilize the behavior of the coherent Higgs field 
after the end of the inflation. However, if the curvature oscillation is very fast, the curvature mass-term 
$\xi (\mu)R\left(t\right)$ cannot generate the exponential growth of the coherent Higgs field $\phi\left( t \right)$
after inflation.

After inflation, the inflaton field $S$ oscillates and 
produces a huge amount of elementary particles.
These particles produced during preheating stage interact with each other and eventually form a thermal plasma.
We comment that thermal effects during reheating stage raise the effective Higgs potential
via the extra effective mass $m^{2}_{\rm eff}=O\left(T^{2}\right)$. 
The one-loop thermal corrections to the Higgs effective potential is given as follows:
\begin{eqnarray}
\Delta { V }_{\rm eff}\left( \phi,T \right)=\sum _{ i=W,Z,t,\phi}{ \frac { { n }_{ i }{ T }^{ 4 } }{ 2{ \pi  }^{ 2 } } \int _{ 0 }^{ \infty  }{ dk{ k }^{ 2 }\ln { \left( 1\mp { e }^{ -\sqrt { { k }^{ 2 }+{{ M }_{ i }^{ 2 }\left( \phi \right)}/{ { T }^{ 2 } }}  } \right)  }    }  } .
\end{eqnarray}
It is possible to lift up the Higgs effective potential throughout the thermal phase via the large thermal effective mass. 
However, it is impossible to prevent 
the exponential growth of the coherent Higgs field $\phi\left( t \right)$ after inflation,
or the large Higgs vacuum fluctuations via the tachyonic resonance during preheating by the thermal effects,
because there can exist a considerable time lag for the production of the thermal bath on the oscillation of the inflaton.
Furthermore, the thermal fluctuations of the Higgs field ${ \left< { \phi  }^{ 2 } \right>  }_{ T }\simeq { T }^{ 2 }/12$ 
might generate the catastrophic AdS domains or bubbles if $T>\Lambda_{I}$ (see e.g. Ref.\cite{Kohri:2016wof,Anderson:1990aa,Arnold:1991cv,Espinosa:1995se,Rose:2015lna} for the details).
Thus, the thermal effects cannot generally guarantee the stability of the electroweak vacuum in the inflationary Universe.
Here, we summarize the conclusions obtained by the above discussion as follows:

\begin{itemize}
\item In $H_{\rm end}> \Lambda_{I} $ and $H_{\rm end}/m_{S}\gtrsim  (\log 10{ \sqrt { 3\xi \left( \mu  \right)  }  })/\xi \left( \mu  \right)$,
the almost coherent Higgs fields  $\phi \left( t \right)$ generated at the end of the inflation
exponentially grow and finally go out to the Planck-energy vacuum state, which leads to the catastrophic collapse of the Universe.
\item In $H_{\rm end}> \Lambda_{I} $ and $H_{\rm end}/m_{S}\gtrsim  (\log { \sqrt { 3\xi \left( \mu  \right)  }  })/\xi \left( \mu  \right)$,
the coherent Higgs field  $\phi \left( t \right)$ on one horizon-size domain
exponentially grows at the end of the inflation and 
forms catastrophic AdS domains or bubbles, which finally cause the vacuum transition of the Universe.
\item In the tachyonic resonance regime $\Phi^{2}\xi \gtrsim M_{\rm pl}^{2}$,
the Higgs field fluctuations extremely increase as 
$\left< {  \delta \phi  }^{ 2 } \right>_{\rm ren}\gg O\left( H^{2}\left( t \right) \right)$. 
Therefore, the effective potential becomes negative i.e. $V_{\rm eff}'\left( \phi  \right)\lesssim 0$, and
the excursion of the homogeneous Higgs field $\phi \left( t \right)$ to the
negative Planck-energy vacuum state occurs during preheating stage and cause the catastrophic collapse of the Universe.
\item In the narrow resonance regime $\Phi^{2}\xi < M_{\rm pl}^{2}$, the Higgs field fluctuations
decrease as $\left< {  \delta \phi  }^{ 2 } \right>_{\rm ren}\simeq O\left( H^{2}\left(t \right)\right)$
due the expansion of the Universe, and therefore, it is improbable to destabilize
the effective potential during preheating stage.
\end{itemize}

The relative large non-minimal Higgs-gravity coupling as $\xi (\mu)\gtrsim O\left(10^{-2}\right)$ can
stabilize the effective Higgs potential and suppress the formations of  the catastrophic AdS domains or bubbles during inflation.
However, after inflation, the effective mass-term $\xi (\mu)R$ via the non-minimal coupling drops rapidly,
sometimes become negative and lead to the exponential growth of the coherent Higgs field $\phi\left( t \right)$ 
at the end of inflation, or the large Higgs vacuum fluctuations via the tachyonic resonance during preheating stage.
Therefore, the non-minimally coupling $\xi(\mu)$ cannot prevent the catastrophic scenario during or after inflation.
After all, if we have large Hubble scale $H> \Lambda_{I} $,
meaning the relatively large tensor-to-scalar ratio $r_{T}$ from 
the polarization measurements of the cosmic microwave background radiation,
the safety of our electroweak vacuum is inevitably threatened during inflation or after inflation 
by the behavior of the homogeneous Higgs field $\phi$ or the generations of the catastrophic AdS domains or bubbles.
We can simply avoid this situation by assuming the inflaton-Higgs couplings $\lambda_{\phi S}$~\cite{Lebedev:2012sy},
the inflationary stabilizations~\cite{Ema:2016ehh,Kawasaki:2016ijp}, or the high-order corrections from GUT or Planck-scale new physics~\cite{Branchina:2013jra,Lalak:2014qua,Branchina:2014efa,Branchina:2014usa,Branchina:2014rva} etc.
In any case, however, the electroweak vacuum instability from inflation gives tight constraints on the beyond the standard model.

\section{Conclusion}
\label{sec:Conclusion}
In this work, we have investigated the electroweak vacuum instability during or after inflation.
In the inflationary Universe, i.e., de Sitter space, the vacuum field fluctuations $\left< {\delta \phi  }^{ 2 } \right>$ 
enlarge in proportion to the Hubble scale $H^{2}$. Therefore, the large inflationary vacuum fluctuations of 
the Higgs field $\left< {\delta \phi  }^{ 2 } \right>$ is potentially catastrophic to trigger the vacuum decay to 
a negative-energy Planck-scale vacuum state and cause an immediate collapse of the Universe. 
However, the vacuum field fluctuations $\left< {\delta \phi  }^{ 2 } \right>$, i.e., 
the vacuum expectation values have a ultraviolet divergence, which is a well-known fact in quantum field theory, 
and therefore a renormalization is necessary to estimate the physical effects of the vacuum transition. 
Thus, we have revisited the electroweak vacuum instability during or after inflation from the legitimate perspective 
of QFT in curved space-time. We have discussed dynamics of homogeneous Higgs field $\phi$ determined
by the effective potential ${ V }_{\rm eff}\left( \phi \right)$ in curved space-time
and  the renormalized vacuum fluctuations $\left< {\delta \phi  }^{ 2 } \right>_{\rm ren}$ 
by using adiabatic regularization and point-splitting regularization, 
where we assumed the simple scenario that the Higgs field only couples the gravity via 
the non-minimal Higgs-gravity coupling $\xi(\mu)$. 
In this scenario, we conclude that the Hubble scale must be smaller than $H<\Lambda_{I} $, 
or the Higgs effective potential in curved space-time 
is stabilized below the Planck scale by a new physics beyond the standard model.
Otherwise, our electroweak vacuum is inevitably threatened by the catastrophic behavior of the homogeneous 
Higgs field $\phi$ or the formations of AdS domains or bubbles during or after inflation.

\acknowledgments
This work is supported in part by MEXT KAKENHI No.15H05889 and No.JP16H00877 (K.K.), and JSPS KAKENHI
Nos.26105520 and 26247042 (K.K.).


\bibliographystyle{JHEP}
\bibliography{renormalization}
\end{document}